\documentclass[pra,superscriptaddress,groupedaddress,twocolumn]{revtex4-1}
\usepackage{amssymb}
\usepackage{amsmath}
\usepackage{graphicx}
\usepackage{dcolumn}
\usepackage{color}
\usepackage{bm}
\usepackage{epsfig}
\usepackage[percent]{overpic}

\newcommand{\white}[1]{\textcolor{white}{#1}}

\DeclareMathOperator{\sech}{sech}

\begin{document}

\title{Mobility of solitons in one-dimensional lattices with the
cubic-quintic nonlinearity}
\author{C. Mej\'ia-Cort\'es}
\email[Corresponding author: ]{ccmejia@googlemail.com}
\author{Rodrigo A. Vicencio}
\affiliation{Departamento de F\'isica and MSI-Nucleus on Advanced
Optics, Center for Optics and Photonics (CEFOP), Facultad de
Ciencias, Universidad de Chile, Santiago, Chile}
\author{Boris A. Malomed}
\affiliation{Department of Physical Electronics, School of
Electrical Engineering,  Faculty of Engineering, Tel Aviv
University,Tel Aviv 69978, Israel}

\date{\today }

\begin{abstract}
We investigate mobility regimes for localized modes in the discrete nonlinear
Schr\"{o}dinger (DNLS) equation with the cubic-quintic onsite terms. Using the
variational approximation (VA), the largest soliton's total power admitting
progressive motion of kicked discrete solitons is predicted, by comparing the
effective kinetic energy with the respective Peierls-Nabarro (PN) potential
barrier. The prediction is novel for the DNLS model with the cubic-only
nonlinearity too, demonstrating a reasonable agreement with numerical findings.
Small self-focusing quintic term quickly suppresses the mobility. In the case
of the competition between the cubic self-focusing and quintic self-defocusing
terms, we identify parameter regions where odd and even fundamental modes
exchange their stability, involving intermediate asymmetric modes. In this
case, stable solitons can be set in motion by kicking, so as to let them pass
the PN barrier. Unstable solitons spontaneously start oscillatory or
progressive motion, if they are located, respectively, below or above a
mobility threshold. Collisions between moving discrete solitons, at the
competing nonlinearities frame, are studied too.
\end{abstract}

\maketitle

\affiliation{Departamento de F\'isica and MSI-Nucleus on Advanced Optics,
Center for Optics and Photonics (CEFOP), Facultad de Ciencias, Universidad de
Chile, Santiago, Chile}

\affiliation{Department of Physical Electronics, School of Electrical
Engineering,  Faculty of Engineering, Tel Aviv University,Tel Aviv 69978,
Israel}

\section{Introduction}

Diffraction of light and matter waves upon propagation is a commonly known
fundamental effect. A wave tends to spread over the whole space as it evolves.
However, when the medium is sensitive to the intensity of traveling waves,
nonlinear corrections must be included in the description the wave propagation,
which often leads to adequate models based on nonlinear partial differential
equations (PDEs). Solitary-wave solutions, or \emph{solitons}, are robust
localized modes generated by this type of nonlinear evolution equations.
Arguably, one of the most generic PDEs related to these systems is the
nonlinear Schr\"{o}dinger (NLS) equation. In the context of optics, it predicts
the existence of effectively one dimensional (1D) solitons in optical
fibers~\cite{Hasegawa2002} and planar waveguides~\cite{Aitchison:91}.

Still richer phenomenology emerges when the nonlinear media include a periodic
transverse modulation of local properties. Nonlinearity and periodicity combine
to offer a wide range of phenomena which have no counterparts in bulk
homogeneous media. Among them, a great deal of interest has been drawn to
discrete
solitons~\cite{fleis,Christodoulides2003,Flach1998181,kevrekidis2009discrete}.
Numerous realizations of discrete solitons have been established, ranging
from nonlinear optics (guided waves in inhomogeneous optical structures
\cite{Eisenberg:02} and photonic-crystal lattices
\cite{PhysRevE.66.046602,1158804}), to atomic physics (chains of droplets of
Bose-Einstein condensates (BEC) trapped in deep spatially periodic
potentials, ~\cite{1367-2630-5-1-371}, \cite{PhysRevA.64.043606},
\cite{PhysRevE.66.046608}, including dipolar BEC with the long-range
interaction between the drops \citep{Sandra1D,Sandra2D}), and from solid-state
settings (Josephson-junction ladders \citep{Matteo,fistul:725,mazo:733}), to
biophysics, in various models of the DNA double strand
\citep{PhysRevE.47.R44,Peyrard1993104}.

Robust mobility of fundamental modes is a necessary ingredient underlying the
transport of light or matter through the lattice. A vast set of theoretical
predictions~\cite{Aceves:94,Vicencio:03,Cuevas} and experimental observations
\cite{PhysRevLett.83.2726,Peschel:02} addressed the issue of the mobility of
discrete solitons in photonic lattices with the cubic (Kerr) nonlinearity, as
well as with saturable \citep{PhysRevE.73.046602} and quadratic
(second-harmonic-generating) \cite{PhysRevLett.99.214103} nonlinear responses.
Recently, it was also analyzed in the model of the photonic lattice with a
saturable nonlinearity~\cite{Naether:11}, and in an array of coupled optical
resonators~\cite{Egorov:13}. Adopting the cubic-quintic (CQ) form of the
on-site nonlinear terms, i.e., including the second-order Kerr corrections to
the refractive index, the solitons have been observed traveling across the
lattice after being precisely kicked~\cite{PhysRevE.77.036604}. As concerns
applications, a system that exhibits good mobility of localized modes may be
promising for the design of all-optical networks, where mobility can help to
implement fast switching and transfer of signals across the system.

In this work we focus on the existence and stability of mobile fundamental
localized modes, both odd and even ones (alias on-site and off-site-centered
states, respectively), in the discrete NLS (DNLS) equation with CQ on-site
nonlinearities. First, we develop an analytical approach predicting, with the
help of the variational approximation (VA), the \textit{mobility threshold} in
this model; i.e., the largest value of the total power (alias norm) of the
discrete soliton admitting its progressive motion induced by the initial kick.
The threshold is predicted by equating the largest possible value if effective
kinetic energy of the kicked soliton to the height of the corresponding
Peierls-Nabarro (PN) barrier. This analytical result is new for the usual DNLS
lattice with the purely cubic nonlinearity too. Its comparison with numerical
results demonstrates a reasonable agreement. The VA-based prediction describes
qualitatively correctly too fast suppression of the mobility of the discrete
solitons with the increase of the self-focusing quintic nonlinear term. In the
case of the competing combination of the self-focusing cubic and defocusing
quintic terms, we emphasize the crucial role played by regions of the stability
exchange between the odd and even fundamental modes, and by intermediate
asymmetric modes existing in those regions, for the understanding of the
\emph{spontaneous mobility} featured by solitons which are unstable in the
static state. We also study the mobility of stable discrete solitons under the
action of the kick.

The rest of the paper is organized as follows. The model is formulated in
Section II. The VA for the discrete solitons and the PN\ barrier are considered
in the analytical form in Section III, the main result of which is the
prediction of the above-mentioned mobility threshold. Numerical results for the
system with the absolute-focusing nonlinearity are reported in Section IV (in
particular, the analytically predicted mobility threshold is compared to
numerical findings). The model with the competing focusing-defocusing CQ terms
is considered, in the numerical form, in Section V. In this section, points of
the stability exchange between the odd and even fundamental localized modes are
analyzed at first, due to the specific role which, as said above, they play in
the transition of unstable solitons into the state of spontaneous motion. The
dynamical regimes proper, both of the spontaneous motion for originally
unstable discrete modes, and the motion of kicked stable solitons, are
presented in that section too, along with basic results for collisions between
moving discrete solitons. The paper is concluded by Section VI.

\section{The model}

We address the propagation of waves in nonlinear discrete systems which were
studied in diverse physical contexts in the course of the last two
decades~\cite{campbell:43,Lederer20081,Flach20081,kevrekidis2009discrete}.  In
those systems, the discreteness appears as an effect of weak interaction
between separated elements, basic examples being arrays of coupled optical
waveguides or BEC trapped in deep optical lattices. In the context of optics,
an evanescent coupling (linear interaction) between modes of adjacent
waveguiding cores takes place when the waveguides are set in close proximity to
each other.  Considering this interaction in the lattice, a set of linearly
coupled equations is derived, similar to the tight-binding models in
solid-state physics \cite{PhysRevB.58.7260}. Assuming a local (on-site)
cubic-quintic nonlinear response of the system, we arrive at the 1D DNLS
equation in the known form:
\begin{equation}
i\dot{\psi _{n}}+(\psi _{n+1}+\psi _{n-1})+\gamma |\psi _{n}|^{2}\psi
_{n}+\nu |\psi _{n}|^{4}\psi _{n}=0\ .  \label{f1}
\end{equation}
Here, $\psi _{n}(z)$ is the field amplitude at the $n$-th lattice site, and
$\dot{\psi}_{n}$ stands for its derivative with respect to the evolutional
variable, which corresponds to the normalized propagation coordinate $z$ in the
present case (this model also applies to BECs loaded into a deep optical
lattice, where the dynamical variable is time $t$).  The inter-site coupling
constant is fixed here to be $1$, but the coefficient in front of the onsite
quintic term is a free parameter, $\nu $.  For $\gamma >0$ and $\nu <0$ (the
competing onsite self-focusing cubic and self-defocusing quintic terms), static
unstaggered solitons in this model were studied in detail in
Ref.~\cite{CarreteroGonzalez200677}, and their 2D counterparts -- in
Ref.~\cite{Chong2009126}.

Equations (\ref{f1}) conserve two dynamical invariants, namely, the total
power, alias norm,
\begin{equation}
P=\sum_{n=-N}^{N}|\psi _{n}|^{2}\ ,  \label{f2}
\end{equation}
and the Hamiltonian
\begin{equation}
H=-\sum_{n=-N}^{N}\Big(\psi _{n+1}\psi _{n}^{\ast }+\psi _{n+1}^{\ast }\psi
_{n}+\frac{\gamma }{2}|\psi |^{4}+\frac{\nu }{3}|\psi |^{6}\Big)\ ,
\label{f3}
\end{equation}
for a lattice of $2N+1$ sites.

First, we look for stationary solutions of Eqs. (\ref{f1}) of the usual form,
$\psi _{n}(z)=\phi _{n}\exp (i\lambda z)$, where $\lambda $ is to the
longitudinal propagation constant, while $\phi _{n}$ defines a real spatial
profile of the soliton. Linear solutions, in the case of $\gamma ,\nu =0$,
correspond to plane waves $\phi _{n}=A\exp (ikn)$ that form the linear band of
the system defined as $\lambda =2\cos k$, where $k$ is the transverse
propagation constant. Thus, the linear band covers the region
$\lambda\in(-2,2)$.

In the nonlinear system, with $\gamma ,\nu \neq 0$, exponentially localized
solutions for discrete solitons exist outside of the linear band, and they
can be obtained by solving the following set of real coupled algebraic
equations:
\begin{equation}
\lambda \phi _{n}=\phi _{n+1}+\phi _{n-1}+\gamma \phi _{n}^{3}+\nu \phi
_{n}^{5}\ .  \label{f5}
\end{equation}
By means of the high-confinement approximation~\cite{Falk_spring}, we construct
an initial condition which is very close to an exact localized solution. Then,
we implement an iterative multidimensional Newton-Raphson method to find a
numerical solution for a given frequency $\lambda $ or power $P$ (both
parameters can be used independently to find solutions).  Once a solution is
obtained, we vary the parameters to construct a whole family of discrete
solitons. Then, we perform the standard linear stability
analysis~\cite{0305-4470-38-4-002}, obtaining an instability growth rate, $g$,
for each solution. In our notation, $g=0$ represents a stable solution, while
$g>0$ an unstable one. Hereafter, stable and unstable solutions are plotted
by means of solid and dashed lines, respectively.

In the framework of this work, we focus on typical fundamental modes that are
relevant to understand the dynamical properties of collective excitations in
this type of lattices. Accordingly, we will call an \textit{odd mode} the one
centered at one site of the lattice, while the \textit{even mode} is centered
between two lattice sites. Other frequently used names for these two species of
the localized states are, respectively, on-site-centered and off-site-centered
modes. The quantity which unequivocally identifies the species is the
center-of-mass coordinate, defined as
\begin{equation}
X_{cm}\equiv \frac{1}{P}\sum_{n=-N}^{N}n|\phi _{n}|^{2}\ .
\end{equation}
Odd and even modes are singled out, severally, by integer and semi-integer
values of $X_{cm}$, respectively.

\section{The energy barrier and variational approximation}

\subsection{The Peierls-Nabarro barrier}

In DNLS models, the PN potential is a major concept for studying the mobility
of localized modes. Odd and even states correspond to fundamental quiescent
solutions. The difference in their Hamiltonian values defines the height of the
PN potential barrier~\cite{PhysRevE.48.3077}.  Indeed, for a given power
(norm), this difference determines effective energetic barriers for moving the
discrete soliton across the lattice: to move the odd mode a single site across
the lattice, it needs to be transformed into even mode and, again, to the odd
mode centered at the next site. Thus, the energy required for this
transformation must be at least equal to the energy difference between the two
states, keeping the power constant (in the case of the adiabatic movement).

It is worthy to mention that different nonlinear interactions generate diverse
energy landscapes. For example, a photorefractive saturable DNLS
model~\cite{PhysRevA.87.043837,Naether:11,PhysRevE.83.036601} allows the
existence of a single stability region for any fundamental solution, and
regions of multi-stability where intermediate asymmetric solutions appear.  The
continuous exchange of stability properties is accompanied by several crossings
of the Hamiltonian values, generating diverse regions where the mobility is
enhanced, even for high powers. Another interesting model is a dipolar DNLS
one, applied for the description of BEC's with long-range
interactions~\cite{PhysRevA.84.033621}. This cubic model possesses a single
region of simultaneously unstable solutions, where the stable solution becomes
the intermediate one. Good mobility occurs in the exchange regions, where
Hamiltonian values approach and cross each other.

In this connection, it is also relevant to mention a very recent result for 2D
cubic Kagome lattices~\cite{PhysRevA.87.061803}. For that model, it was found
that a single stability-exchange region appears, where the intermediate
solution becomes the ground state of the system. As the appearance of these
solutions implies that the energy barriers decrease, coherent transport for
very low power and highly localized modes were also found, what is an unusual
property of 2D cubic
lattices~\cite{campbell:43,Lederer20081,Flach20081,kevrekidis2009discrete}.

\subsection{Variational approximation}

The variational approximation (VA) provides a first estimate about regions
in the parameter space, where mobility may be expected. The VA was developed in
Refs.~\cite{PhysRevE.77.036604} and \cite{CarreteroGonzalez200677} for 1D
discrete solitons in the CQ-DNLS model, but considering only the
self-defocusing sign of the quintic term. To apply the VA, the following
\textit{ansatz} was adopted:
\begin{equation}
\phi _{n}^{(\mathrm{VA)}}=Ae^{-\alpha |n-n_{0}|}\ ,  \label{trial}
\end{equation}
where amplitude $A$ and inverse width $\alpha $ are real positive constants,
and $n_{0}=(\chi +1)/2$ defines the position of the center of the mode, with
$\chi =0$ and $\chi =1$ for even and odd states, respectively. We here treat
$A$ as a variational parameter, while $\alpha $ is fixed by the substitution of
ansatz (\ref{trial}) into the linearization of Eq. (\ref{f5}), for the decaying
tail of the discrete soliton far from its center: $\lambda=2\cosh{\alpha}$.

The power associated with ansatz~(\ref{trial}) is
\begin{equation}
P=A^{2}\frac{\cosh \left( \chi \alpha \right)}{\sinh \alpha}.
\label{norm}
\end{equation}

It is well known that the Hamiltonian can be expressed as the Legendre
transform of its Lagrangian, namely,
\begin{equation}
H(q_{n},p_{n})=\sum_{n=-\infty }^{+\infty }{\dot{q}_{n}p_{n}}-L(q_{n},\dot{q}%
_{n})\ ,
\end{equation}
where, for model (\ref{f1}), canonical coordinates $p_{n}$ and $q_{n}$
correspond to $\psi _{n}$ and $i\psi _{n}^{\ast }$, respectively. Thus,
Eq.~(\ref{f5}) corresponds to the following Lagrangian,
\begin{equation}
-L=\lambda P+H.
\label{L}
\end{equation}
The substitution of ansatz (\ref{trial}) into Eq.~(\ref{f3}) and the
elimination of $A^{2}$ by means of Eq. (\ref{norm}) yields the effective
Hamiltonians for the odd and even modes,
\begin{widetext}
\begin{gather}
H^{\mathrm{VA}}_{\rm {odd}}(P)=-2\sech{(\alpha)}P
-\frac{\gamma}{4}\left[\cosh(2\alpha)\sech^2(\alpha)\tanh(\alpha)\right]P^2
-\frac{\nu}{3}\left[\frac{2\cosh(2\alpha) -
1}{2\cosh(2\alpha)+1}\tanh^2(\alpha)\right]P^3\ , \label{heffodd}
\end{gather}
\begin{gather}
H^{\mathrm{VA}}_{{\rm
even}}(P)=-2e^{-\alpha}\left[1+\sinh(\alpha)\right]P
-\frac{\gamma}{4}\tanh(\alpha)P^2-
\frac{\nu}{3}\left[\frac{\sinh^2(\alpha)}{2\cosh(2\alpha)+1}\right]P^3\
. \label{heffeven}
\end{gather}
\end{widetext}

Finally, by inserting the previous expressions for the Hamiltonian into
Eq.~(\ref{L}), we obtain the full Lagrangian of the system for both fundamental
modes as function of their width and power content:
\begin{widetext}
\begin{equation}
L_{\mathrm{odd}}^{\mathrm{VA}}(P)=-2\sinh (\alpha )\tanh (\alpha )P
+\frac{\gamma }{4}\left[ \cosh (2\alpha )\sech^{2}(\alpha )\tanh (\alpha )
\right] P^{2}
+\frac{\nu}{3}\ \left[ \frac{2\cosh (2\alpha )-1}{2\cosh (2\alpha )+1}\tanh ^{2}(\alpha
)\right] P^{3}\ ,  \label{leff}
\end{equation}
\begin{equation}
L_{\mathrm{even}}^{\mathrm{VA}}(P)=\left[ 1-\cosh (2\alpha )-2\sinh (\alpha
)+\sinh (2\alpha )\right] P
+\frac{\gamma }{4}\tanh (\alpha )P^{2}+\frac{\nu}{3}\ \left[ \frac{\sinh ^{2}(\alpha )}{
1+2\cosh (2\alpha )}\right] P^{3}.
\end{equation}
\end{widetext}

The Euler-Lagrange equations obtained by these Lagrangians are
\begin{equation}
\frac{\partial }{\partial P}\left[L_{\mathrm{odd,even}}^{(\mathrm{VA)}}\right]=0.
\label{d/d}
\end{equation}
To understand how the quintic term in Eq. (\ref{f1}) affects the mobility of
localized solutions, it may be treated as a perturbation because, as shown
below, quite small positive values of $\nu $ lead to full suppression of the
mobility. Then, in the zero-order approximation ($\nu =0$), Eqs.~(\ref{d/d})
and (\ref{leff}), for the odd modes and $\gamma =1$, yield
\begin{equation}
P_{\mathrm{odd}}^{(0)}(\alpha )=4\left( \cosh ^{2}\alpha \right) \left(
\sinh \alpha \right) \sech(2\alpha )\ .
\label{P0}
\end{equation}
Next, we introduce the first-order correction to Eq.~(\ref{P0}),
\begin{equation}
P_{\mathrm{odd}}(\alpha )=P_{\mathrm{odd}}^{(0)}(\alpha )+\nu P_{\mathrm{odd}%
}^{(1)}(\alpha ),
\label{PPP}
\end{equation}
for which the calculation based on the VA results yields
\begin{equation}
P_{\mathrm{odd}}^{(1)}(\alpha )=32\frac{\left[ 2-\sech(2\alpha
)\right] \left( \cosh ^{5}\alpha \right) \sinh ^{3}\alpha }{\left[
2\cosh (2\alpha ) +1\right] \cosh ^{2}\left( 2\alpha \right) }.
\label{P1odd}
\end{equation}

To estimate the mobility threshold, it is necessary to find the \emph{largest}
values of $\alpha$ and $P$ for which the kicked mode may be set in
motion~\cite{Duncan19931,Flach199961,PhysRevE.59.6105,PhysRevE.65.026602,PhysRevE.68.046604,kevrekidis2009discrete}.
A kicked odd profile is obtained by adding the phase term to ansatz
(\ref{trial}),
\begin{equation}
\phi _{n}^{(0)}\cdot e^{ikn}=Ae^{-\alpha |n|+ikn},
\label{kicked}
\end{equation}
where $k$ is the real magnitude of the kick, which corresponds to a transverse
propagation constant restricted to the interval of $[0,\pi]$.  This kicked
profile increases the effective Hamiltonian of the odd mode by adding what, in
the present context, plays the role of the \emph{kinetic} \emph{energy}:
\begin{equation}
E_{\mathrm{kin}}=2\sech\alpha\left( 1-\cos k\right) P.
\label{Ekin}
\end{equation}

Then, taking into regard that the Hamiltonian of the even mode is larger than
for the odd one, the mobility limit is determined by equating Hamiltonian
(\ref{heffodd}), modified according to ansatz (\ref{kicked}) with $k=\pi $
[which implies taking the largest possible value of kinetic energy
(\ref{Ekin})], to the energy of the immobile even soliton, for the same total
power $P$:
\begin{equation}
H_{\mathrm{odd}}^{(\mathrm{VA)}}(P,k=\pi)=H_{\mathrm{even}}^{(\mathrm{VA)}}(P). 
\label{H=H}
\end{equation}
This equations determines the largest values of the total power, $P_{\max }$,
at which the kicked discrete soliton may be set in persistent motion.

First, in the cubic-only model ($\nu =0$), Eq. (\ref{H=H}), with
$H_{\mathrm{odd,even}}^{(\mathrm{VA)}}$ and $P$ substituted by the VA results
[see Eqs. (\ref{heffodd}), (\ref{heffeven}), and (\ref{P0})], leads to the
following equation for $\alpha _{\max }$, which corresponds to the respective
mobility threshold $P_{\max }^{(0)}$:
\begin{widetext}
\begin{equation}
2\left( 2+\sinh \alpha _{\max }+\tanh \alpha _{\max }\right) \left( \cosh
\alpha _{\max }\right) \cosh \left( 2\alpha _{\max }\right) =\exp{(\alpha _{\max })}\sinh ^{4}\alpha _{\max },
\label{dh}
\end{equation}
\end{widetext}
a numerical solution of which yields
\begin{equation}
\alpha _{\max }\approx 2.0361.
\label{alpha}
\end{equation}
Finally, the substitution of this $\alpha _{\max }$ into the VA-predicted
expression for the total power [see Eq. (\ref{P0})] gives
\begin{equation}
\ P_{\max }^{(0)}=P_{\mathrm{odd}}^{(0)}(\alpha _{\max })\approx 7.7866\ .
\label{Pmax0}
\end{equation}

We stress that this prediction of the mobility threshold, i.e., the largest
total power of the odd discrete soliton which may be set into motion by the
arbitrarily large kick, is a novel result even for the usual DNLS model with
the purely cubic nonlinearity.

Finally, the weak quintic term, if treated as a small perturbation (see
above), shifts the largest value of the total power, which admits the
progressive motion, as
\begin{equation}
P_{\max }\approx P_{\max }^{(0)}+\nu P_{\max }^{(1)}\ ,
\label{P1max}
\end{equation}
with the coefficient found from the respective expansion of Eq.~(\ref{H=H}):
\begin{equation}
P_{\max }^{(1)}\approx \frac{2\sinh ^{4}(\alpha _{\max })\tanh (\alpha
_{\max })}{\sinh (2\alpha _{\max })+\sinh (4\alpha _{\max })}(P_{\max
}^{(0)})^{2}\
\label{p1max}
\end{equation}
[recall $\alpha _{\max }$ is given by Eq.~(\ref{alpha})]. At the same order
$\sim \nu $, it is also necessary to include the correction to the VA-predicted
total power, as given by Eq.~(\ref{PPP}). Thus, taking into regard both
contributions, one from Eq.~(\ref{p1max}) and one from (\ref{P1odd}) with
$\alpha =\alpha _{\max }$, our estimate for the largest total power admitting
the mobility of the discrete soliton in Eq.~(\ref{f1}), including the quintic
term, is
\begin{gather}
P_{\max }=P_{\max }^{(0)}-\nu \left[ P_{\mathrm{odd}}^{(1)}\left( \alpha
_{\max }\right) +P_{\max }^{(1)}\right] \notag\\
\approx 7.7866-72.0148\nu \ .
\label{P1}
\end{gather}
Eventually, the absence of the mobility, alias ``full stop" ($P_{\max }=0$),
is predicted to occur at the following positive value of the quintic
coefficient:
\begin{equation}
\nu _{\mathrm{halt}}=0.1081\ .
\label{halt}
\end{equation}%
On the other hand, Eq. (\ref{P1}) predicts that a negative quintic
coefficient will produce the opposite effect, allowing mobility at larger
values of power.

Comparing these predictions with numerical findings (see below), it is
necessary to take into account that sufficiently heavy kicked localized
modes first shed off a part of their power in the form of the lattice
radiation waves (``phonons"). For this reason, the effective power which is
predicted by Eqs.~(\ref{Pmax0})-(\ref{P1}) is expected to be smaller than
the actual initial power of the kicked modes. It will be seen below that
this expectation is borne out by the comparison with numerical results.

\section{Numerical results for the model with self-focusing cubic and
quintic terms: $\protect\gamma, \protect\nu >0$\ .}

\subsection{Stationary solutions}

\begin{figure}[tb]
\centering
\epsfig{file=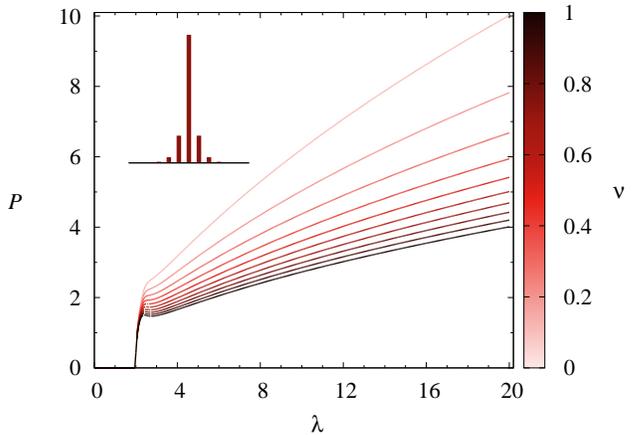,width=1\linewidth,clip=}
\caption{(Color online) Total power $P$ versus propagation constant
$\protect\lambda $ for odd-mode (on-site-centered) soliton families. Each color
corresponds to a different value of quintic coefficient $\protect\nu $, while
the cubic coefficient is fixed to $\protect\gamma =1$. The inset shows a
typical profile of the odd localized mode for $P=2.4$, $\lambda=5.0$ and
$\nu=0.5$.}
\label{fig1}
\end{figure}

\begin{figure}[tb]
\centering
\epsfig{file=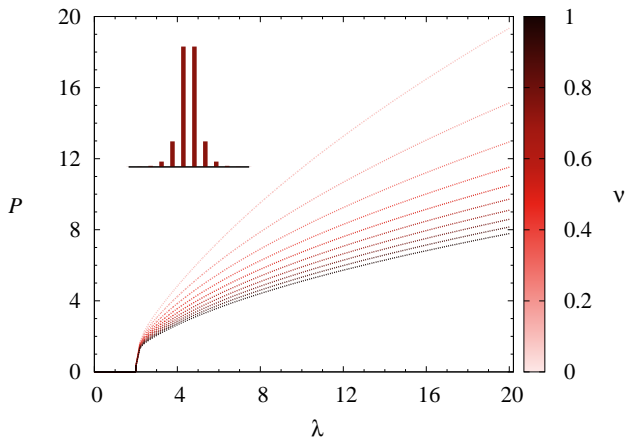,width=1\linewidth,clip=}
\caption{(Color online) The same as in Fig. \protect\ref{fig1}, but for the
families of even (inter-site-centered) modes. The inset shows a typical profile
of the even localized mode for $P=4$, $\lambda=5.0$ and $\nu=0.5$.}
\label{fig2}
\end{figure}

We start the numerical analysis by considering positive values of both
nonlinear coefficients in Eq. (\ref{f1}): $\gamma ,\nu >0$, which corresponds
to both onsite nonlinear terms being \textit{self-focusing}, in the optical
context. First, we construct families of odd and even fundamental localized
modes by solving Eq. (\ref{f2}), for propagation constants above the linear
band. In Figs.~\ref{fig1} and \ref{fig2} we show, different, families of odd
and even fundamental solutions; insets show typical profiles of the odd and
even fundamental modes. In general, with the increase of the quintic
coefficient $\nu$ at constant propagation constant $\lambda$, we observe that
the total power of the localized solutions decreases.  Therefore, under the
stronger self-focusing nonlinearity, it may be possible to excite strongly
localized states at lower levels of power.

As we mentioned above, the stability of these solutions was predicted by means
of a standard linear analysis; additionally it was corroborated by direct
simulations of their evolution in the framework of Eq.~(\ref{f1}), adding
initial white-noise perturbations. The odd mode is found to be unstable only
in extremely small portions of the $P(\lambda )$ curves with the negative slope
$\partial P/\partial \lambda <0$, in agreement with the Vakhitov-Kolokolov (VK)
stability criterion \cite{VKcrit}. On the other hand, the even mode is unstable
in the whole range of explored parameters, which is a typical situation for
cubic DNLS models~\cite{kevrekidis2009discrete}. In the course of the perturbed
evolution, the stable odd modes radiate a very small amount of their power
content and relax into the unperturbed shape. The unstable even modes radiate a
significant amount of power, and oscillate between the odd and even profiles.

\begin{figure}[tb]
\centering
\epsfig{file=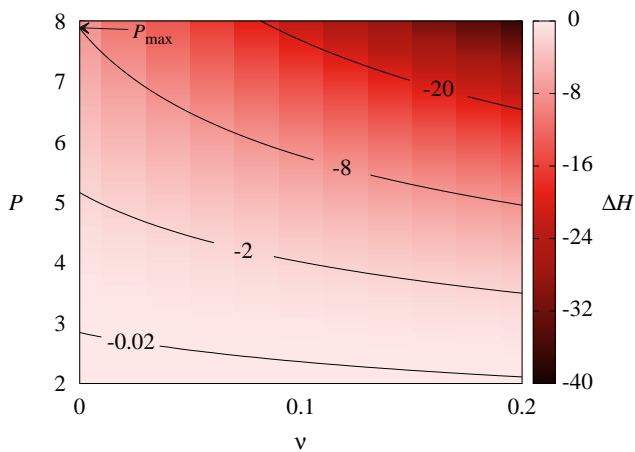,width=1.02\linewidth,clip=}
\caption{(Color online) The color map of the PN barrier, $\Delta H$, versus $ P
$ and $\protect\nu $. The black arrow point out the VA of the
$P_{\mathrm{max}}$ value.}
\label{fig3}
\end{figure}

To study the mobility of the localized modes, we first compute the PN barrier,
which was defined above as the difference between the values of the Hamiltonian
for the odd and the even modes~\cite{PhysRevE.48.3077}: $\Delta H\equiv
H_{\mathrm{odd}}-H_{\mathrm{even}}$. Figure~\ref{fig3} displays a color map
showing that the PN barrier monotonically increases with the total power and
the positive quintic coefficient, $\nu$. Therefore, in this parameter regime,
the transport across the lattice is not expected to be enhanced.

It is relevant to point out here good agreement between the VA prediction and
the numerical calculations. Indeed, it follows from Eqs.~(\ref{Ekin}),
(\ref{alpha}), and (\ref{Pmax0}) that the largest kinetic energy which may be
lent to the odd mode, at $\nu=0$, is
\begin{equation}
E_{\mathrm{kin}}\approx 7.9954. \label{prediction}
\end{equation}
According to the prediction of the VA, this extra energy allows one to equalize
the Hamiltonians of the odd and even modes, i.e., it must be equal to the PN
barrier at the VA-predicted value of the total power, given by
Eq.~(\ref{Pmax0}). The black arrow in Fig.~\ref{fig3} indicates this point,
$P=P_{\mathrm{max}}^{0},\nu=0$, which belongs to the contour level of $\Delta
H=-8$. The latter value is remarkably close to the VA prediction given by
Eq.~(\ref{prediction}).

\subsection{Dynamics}

\begin{figure}[tb]
\centering
\begin{overpic}[width=1\columnwidth]{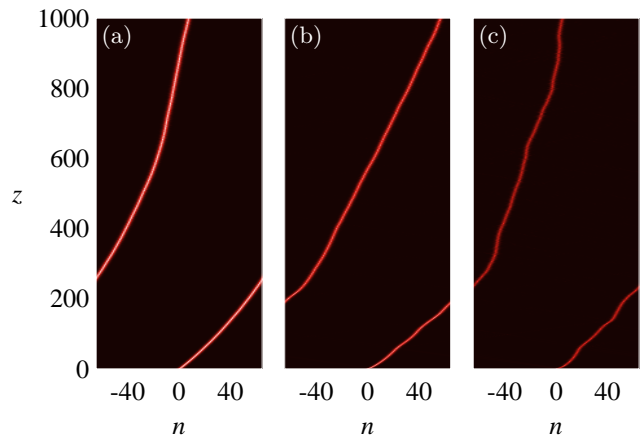}
\put(16,64){\white{(a)}}
\put(45,64){\white{(b)}}
\put(74,64){\white{(c)}}
\end{overpic}
\caption{(Color online) Examples of the mobility of discrete solitons across
the whole lattice with $\protect\gamma =1$ and $\protect\nu =0$ (no quintic
nonlinearity): (a) $P=2.4$ and $k=0.2$, (b) $P=2.8$ and $k=0.5$ and (c) $P=3.2$
and $k=0.8$.}
\label{fig4}
\end{figure}

First, we consider the propagation of localized solutions in the system with
the cubic-only nonlinearity, i.e., $\gamma =1$ and $\nu =0$ in Eq.~(\ref{f1}).
Figure~\ref{fig4} displays examples of clear mobility for three different
powers and three different kicks: (a) $P=2.4$ and $k=0.2$, (b) $P=2.8$ and
$k=0.5$ and (c) $P=3.2$ and $k=0.8$, in lattice with periodic boundary
conditions. It is worthy to note that the trajectories gradually lose their
smoothness as the power increases. Obviously, with the increase of the PN
barrier the mobility deteriorates---in particular, due to the increment in the
emission of radiation. Independent of the size of the kick, stationary modes with
\begin{equation}
P>\left( P_{\max }^{(0)}\right) _{\mathrm{num}}\approx 6
\label{P>6}
\end{equation}
stay immobile, losing a conspicuous part of their power in the form of
radiation as a result of the application of the kick.

The comparison of the numerically found mobility threshold (\ref{P>6}) with the
VA-predicted counterpart (\ref{Pmax0}) demonstrates a sufficiently reasonable
agreement. Taking into regard that the VA for the moving discrete solitons
cannot be developed in a highly accurate form, cf.
Ref.~\cite{PhysRevE.68.046604}. The fact that the predicted threshold value of
the power is greater than the numerically identified one is explained by the
decrease of the power due to the radiative loss, as discussed above.

\begin{figure}[tb]
\centering
\begin{overpic}[width=1\columnwidth]{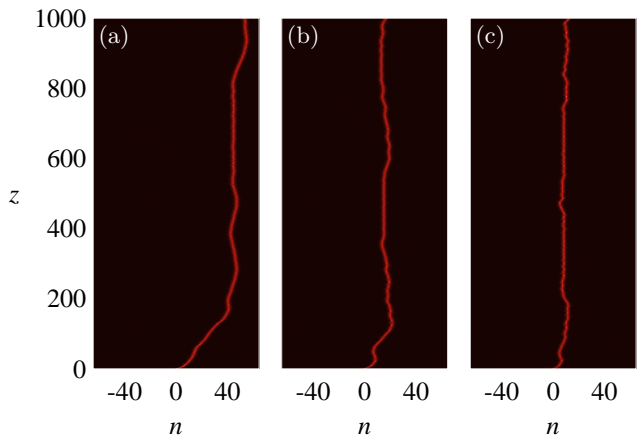}
\put(16,64){\white{(a)}}
\put(45,64){\white{(b)}}
\put(74,64){\white{(c)}}
\end{overpic}
\caption{(Color online) The mobility of discrete solitons with $P=3.2$ under
small values of the quintic coefficient: (a) $\protect\nu =0.01$, (b) $%
\protect\nu =0.03$, and (c) $\protect\nu =0.05$.}
\label{fig5}
\end{figure}

Now, we include the effect of the quintic self-focusing nonlinearity, $\nu>0 $.
As an example, we here take the same values $P=3.2$ and $k=0.8$ as in
Fig.~\ref{fig4}(c). As shown above in Fig.~\ref{fig3} and predicted by
Eq.~(\ref{P1}), the PN barrier increases monotonically with $\nu $, hence
reduction of the mobility is expected. Figure~\ref{fig5} corroborates the
dramatic reduction of the mobility while weakly increasing $\nu $. At $\nu
=0.01 $, the localized mode passes about $40$ sites of the lattice
[Fig.~\ref{fig5}(a)], before getting trapped around $n=41$. Increasing the
quintic coefficient to $\nu =0.03$ and then to $\nu =0.05$, we observe that the
discrete soliton shows erratic motion (it is explained by overcoming the
potential barriers, loosing the energy through the radiation loss, and facing
new effectively lower barriers), before getting trapped around $n=15$
[Fig.~\ref{fig5}(b)] and $n=10$ [Fig.~\ref{fig5}(c)], respectively. Thus, the
prediction of the VA that the quintic self-focusing term must completely
suppress the mobility at rather small values of $\nu $, see Eq.~(\ref{halt}),
is in qualitative agreement with the numerical findings.

\section{Numerical results for the model with self-focusing cubic and
defocusing quintic terms: $\protect\gamma >0,\protect\nu <0$.}

\subsection{Stationary solutions}

\begin{figure}[tb]
\centering \epsfig{file=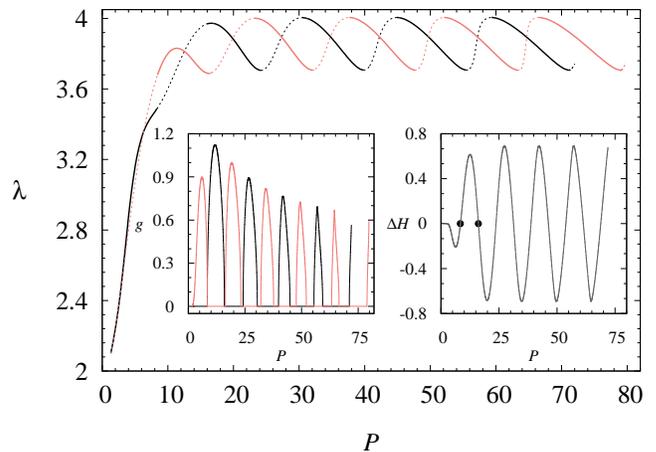,width=1.05\linewidth,clip=}
\caption{(Color online) The propagation constant, $\protect\lambda$, versus the
total power, $P$, for families of odd (black) and even (red) discrete solitons
in the system with $\protect\gamma =1$ and $\protect\nu =-0.1$. Recall that
continuous and dashed curves designate stable and unstable soliton families,
respectively. The left inset shows the instability growth rate, $g$, for these
families. The right inset show the PN barrier corresponding to these solutions,
in the whole domain.}~\label{fig6}
\end{figure}

\begin{figure}[tb]
\epsfig{file=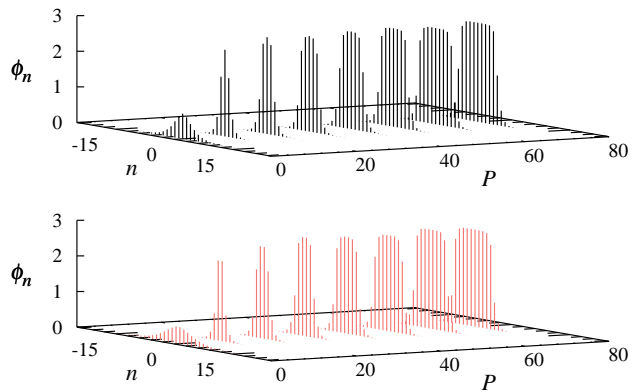,width=1\linewidth,clip=}
\caption{(Color online) Solution profiles for the odd and even modes at several
values of the total power (top and bottom panels, respectively) corresponding
to the families represented by the black and red lines in
Fig.~\protect\ref{fig6}.}
\label{fig7}
\end{figure}

The negative quintic coefficient, $\nu<0$, implies saturation of the on-site
self-focusing nonlinearity (as featured, e.g., by the photorefractive
nonlinearity \cite{PhysRevE.66.046602}). First, we look for fundamental
stationary localized modes in this case. The respective solution families are
constructed by fixing the total power and numerically looking for the
corresponding mode profiles $\phi _{n}$ and the propagation constant $\lambda$,
see Eq.~(\ref{f5}). The results are displayed in Fig.~\ref{fig6} for $\gamma
=1$ and $\nu =-0.1$. We observe that, with the increase of the total power, the
$\lambda (P)$ curves oscillate,  odd (black) and even (red) modes, while the
modal profiles increase their width by adding new sites in each oscillation,
as shown in Fig.~\ref{fig7}.

Below the first crossing point ($\lambda \approx 3.4$), the VK stability
criterion applies. Then, the oscillations give rise to multiple stability
exchanges at several crossing points. By increasing the total power, we observe
regions where both, odd and even, solutions are unstable simultaneously,
regions where only one solutions is stable, and also regions where both
solutions are \emph{simultaneously stable}. The stability switching occurs, as
it is generic for nonlinear dynamical
systems~\cite{burke:037102,Herring2005144,PhysRevLett.103.194101}, at
saddle-node bifurcation points. In this case, these points coincide with maxima
and minima values $\lambda (P)$ for $P>15$. This scenario is similar to the one
known from the saturable nonlinearities and it is characterized by a continuous
spatial broadening of the fundamental solutions, as a consequence of the
saturation of the amplitude at large values of $P$, see Fig.~\ref{fig7}.

\begin{figure}[tb]
\epsfig{file=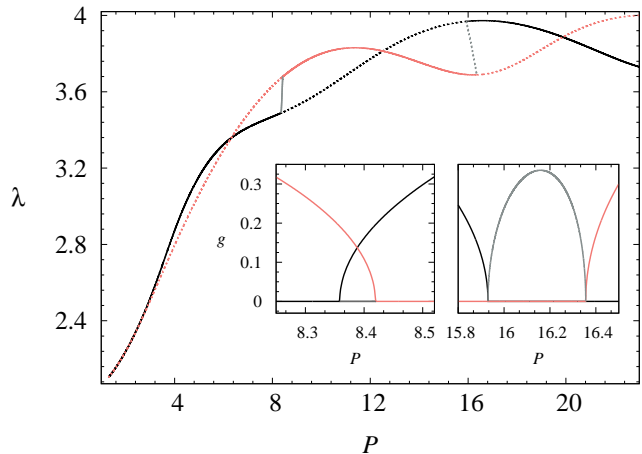,width=1.05\linewidth,clip=}
\caption{(Color online) A zoom of the first stability-exchange region from
Fig.~\protect\ref{fig6}. Asymmetric stable solutions family (the gray curves)
appears as the connection between the odd- and even-mode families (black and
red curves). The inset shows the instability growth rate, $g$, associated with
the solutions belonging to the three different families.} 
\label{fig8}
\end{figure}

On the other hand, by plotting the PN barrier, $\Delta H$, right inset at
Fig.~\ref{fig6}, we observe several points at which it exactly vanishes,
indicating that both solutions share their properties in the Hamiltonian
representation; i.e., both are maxima or minima (unstable and stable states,
respectively).  A straightforward assumption was that at these points the
system becomes ``transparent"~\cite{PhysRevLett.93.033901}, featuring perfect
mobility.  However, a key ingredient was missing: when (at least) two solutions
share stability properties, an extra solution in between must appear with the
opposite stability. These intermediate stationary solutions typically
correspond to asymmetric profiles. In the present case, for the competing CQ
onsite nonlinearity, they were first found in
Ref.~\cite{CarreteroGonzalez200677}. With this ingredient included, the
effective energy barrier is the one that takes into regard \emph{all}
stationary solutions in a certain region of the parameter space, rather than
just the usual odd and even modes. This situation can be understood from the
segment of $\lambda (P)$ curve in Fig.~\ref{fig6} which is zoomed in
Fig.~\ref{fig8}. In this case, families of asymmetric solutions (gray branches)
emerge, linking the odd and even families exactly at the stability-exchange
points. They are associated with the two first vanishing points of the PN
barrier, designated by the black filled circles in Fig.~\ref{fig6}.  Within a
small total-power interval around $P\approx 8.4$, the odd and even families are
simultaneously unstable. On the contrary, around $P=16.1$ both of them are
simultaneously stable. As the power increases, the PN barrier vanishes at
several points, in accordance with oscillation of the $\lambda (P)$ curves and
the exchange of the stability properties.

\begin{figure}[tb]
\epsfig{file=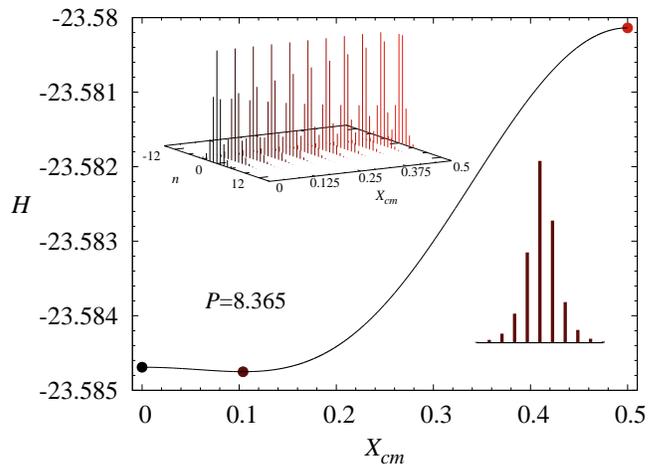,width=1.05\linewidth,clip=}
\caption{(Color online) The $H$ versus $X_{\mathrm{cm}}$ plot corresponding to
the adiabatic transition between the odd- and even-mode solutions. The insets
display asymmetric profiles of the intermediate solutions, see the text.}
\label{fig9}
\end{figure}

There are, essentially, two ways of obtaining the intermediate asymmetric
solutions. The first one amounts to finding this solution dynamically, by
tracing the center of mass of the wave packet and noticing when the velocity is
smaller or larger, corresponding to a maxima or minima of the effective
potential and, accordingly, to unstable or stable
solutions~\cite{PhysRevE.73.046602,PhysRevA.87.061803}. The second, and more
elegant, procedure is provided by the so-called ``\textit{constraint method}"
\cite{Molina:06,PhysRevLett.97.083901}. It starts from a given stationary mode
with defined propagation constant $\lambda $, power $P$, and center-of-mass
coordinate $X$. We then implement a constrained Newton-Raphson method that
finds solutions by keeping the power constant and varying $X$. By performing a
smooth sweep in $X$, we are able to transform the profile from a given
fundamental mode to the other one (e.g., from an odd mode to an even one).
In Fig.~\ref{fig9} we show the $H(X)$ dependence obtained by implementing this
method, including mode profiles of the constrained solutions. In fact, this
process corresponds to a correct definition of the PN potential, the one which
traces the shape of the effective potential that the mode must overcome while
it moves adiabatically (without emission of radiation and keeping the power
constant) from one site to another. As the system is periodic, this potential
must be copied to construct the whole effective lattice potential. If there
exists any intermediate stationary solution, it will correspond to a critical
point on this diagram, depending on the particular model. The filled circles in
Fig.~\ref{fig9} correspond to the stationary fundamental solutions for this
level of power. The black and light red ones, located at $X=0$ and $X=0.5$,
correspond to the unstable odd and even mode solutions. The extra critical
point, the dark red filled circle located in between the other two ones,
corresponds to a stable intermediate solution (potential minimum), with a
characteristic profile sketched in the inset of Fig.~\ref{fig9}.

When finding such an asymmetric mode, we construct the whole family by using a
normal Newton-Raphson method. In Fig.~\ref{fig8} this family is shown by a gray
line, connecting the fundamental odd and even modes in the region of
multi-instability. As this solution corresponds to a minimum of the Hamiltonian
(the ground state for this level total power), it is a stable solution located
between two unstable ones~\cite{arnol1989mathematical}.

\begin{figure}[tb]
\begin{overpic}[width=1 \columnwidth]{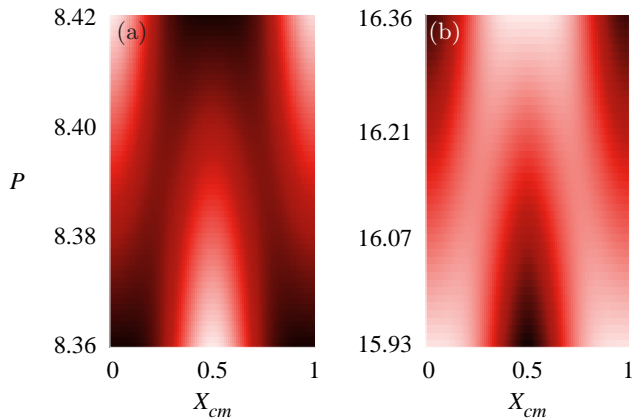}
\put(18,62){(a)}
\put(66,62){\white{(b)}}
\end{overpic}
\caption{(Color online) The color-map plot of the normalized Hamiltonian ($H$)
for the intermediate solutions in the vicinity of stability-exchange points.
(a) \textit{bi-unstable} region and (b) \textit{bi-stable} region.}
\label{fig10}
\end{figure}

We have computed the effective potential for the first \textit{bi-unstable}
region, by sweeping the value of the total power, $P$, and finding the
Hamiltonian for all the constrained solutions in between the odd ($X=0$) and
even ($X=0.5$) modes. To compare different effective potentials, we
normalized the Hamiltonian to its maximum value for any level of the total
power. Figure~\ref{fig10}(a) displays a surface plot of the normalized values
of $H$ as a function of $P$ and $X$. The center of mass of the stable
asymmetric solutions passes from the odd mode to the even one, following the
increase of $P$. For low levels of the total power the odd mode is stable
and the even one unstable. Then, both solutions become unstable up to a
level of the total power at which the even mode transforms into the ground
state, and the odd mode into an unstable solution. In that sense, the
asymmetric intermediate solution plays the role of a \textit{stability
carrier}, which transfers the stability between the fundamental stationary
solutions.

Next, we construct a similar energy landscape in a region where both
fundamental solutions are simultaneously stable, around $P\approx 16.2$.  This
implies the existence of an intermediate asymmetric unstable solution
corresponding to a maximum between two minima (odd and even modes). The
surface plot of the effective potential for this stability region, produced by
the constraint method, is displayed in Fig.~\ref{fig10}(b), which demonstrates
that the intermediate solution is indeed unstable, and, in this region, it
transfers the instability from one fundamental mode to the other, completely
switching their stability properties.

\subsection{Moving solitons}

\begin{figure}[tb]
\epsfig{file=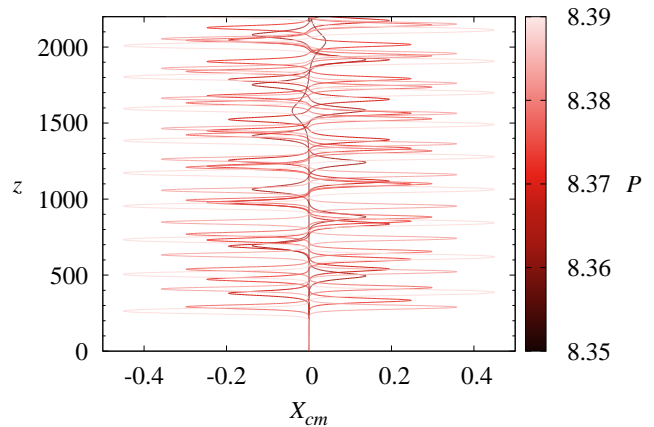,width=1\linewidth,clip=}
\caption{(Color online) Oscillations of the center of mass ($X_{\mathrm{cm}}$)
of unstable odd-mode solutions in the case of $P<P_{\mathrm{th}}$.  Increase of
the total power leads to larger oscillations amplitudes.}
\label{fig11}
\end{figure}

Having established the maps of the effective potential, we here address the
mobility of the localized modes for powers above and below the threshold value
$P_{\mathrm{th}}$, at which $\Delta H(P_{\mathrm{th}})=0$. We start by
analyzing the evolution of unstable odd modes in the bi-unstable region, for a
power below $P_{\mathrm{th}}\approx 8.39$ (see Fig.~\ref{fig10}, where the
value of $H$ for the odd mode increases up to coinciding with that for the even
mode). Below this power level, the odd modes do not have enough energy to
overcome the PN barrier. Thus, the unstable odd modes oscillate in the interval
of $-0.5<X<+0.5$ without being able to cross the barrier.  Figure~\ref{fig11}
displays the evolution of $X$ for several unstable odd modes, all with
$P<P_{\mathrm{th}}$. As the instability growth rate $g$ increases, following
the increase of the total power, the amplitude of the spontaneous oscillations
increases too, because the value of $H$ for the odd mode approaches that for
the even one.

\begin{figure}[tb]
\begin{overpic}[width=1 \columnwidth]{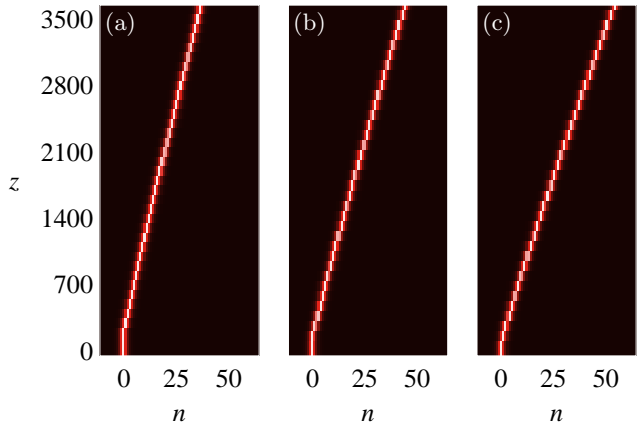}
\put(17,64){\white{(a)}}
\put(46,64){\white{(b)}}
\put(75,64){\white{(c)}}
\end{overpic}
\caption{(Color online) Dynamics of unstable odd-mode solutions for
$P>P_{\mathrm{th}}$. (a) $P=8.92$, (b) $P=8.404$ and (c) $P=8.414$.}
\label{fig12}
\end{figure}

On the other hand, we observe mobility of the localized modes at $P>P_{%
\mathrm{th}}$, where the value of $H$ for the odd mode is larger than the even
one. Therefore, the odd mode is able to overcome the PN barrier and move across
the lattice. Figure~\ref{fig12} shows the evolution of three different
unstable odd-mode solutions for three different levels of the total power. We
observe that the transversal velocity of the spontaneous motion is correlated
to the power; for example, in case (c) the velocity almost doubles in
comparison with case (a). For an increasing power, the value of $H$ for the odd
mode increases in comparison to the even mode. Therefore, the kinetic energy
for this mode is larger, letting it move faster across the lattice.

\begin{figure}[tb]
\epsfig{file=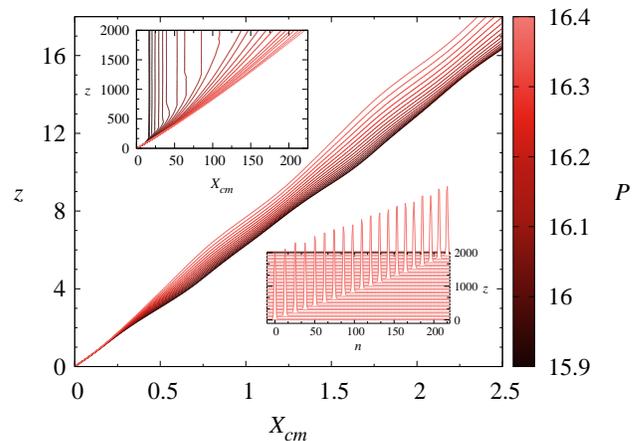,width=1\linewidth,clip=}
\caption{(Color online) The initial stage of the motion of the center of mass,
$X_{\mathrm{cm}}$, of stable odd-mode solutions, with $15.93<P<16.36$, under
the action of kick $k=0.1$. The top inset shows long-scale evolution of these
modes. The bottom inset illustrates the robust mobility of the mode with
$P=16.36$.}
\label{fig13}
\end{figure}

We proceed to exploring the next stability-exchange region, located around
$P\approx 16.14$. Here, both fundamental solutions are stable and both
correspond to minima of the Hamiltonian, without starting spontaneous motion.
Therefore, a kick must be applied to these modes to help them overcome the
energy barriers. The kick with $k=0.1$ was enough to set into motion all the
odd stable modes in this region.  Figure~\ref{fig13} displays the motion of the
center of mass, $X$, for several kicked odd modes with different total powers.
After passing a certain distance, some the kicked discrete solitons come to a
halt (see the upper inset), getting trapped between two sites of the lattice,
while others continue the persistent motion. We observe that the final values
of $X$, at which the motion ceases, increases with the total power of the
soliton, in this region of parameters. On the other hand, Fig.~\ref{fig13} also
shows that the discrete solitons with  smaller total powers initially move
faster than ones with greater powers. This feature can be easily explained by
the comparison with the NLS equation in continuum, where the kick measures the
momentum imparted to the soliton, while the total power determines its mass,
hence the heavier soliton moves slower, under the action of the same kick.

\begin{figure}[tb]
\begin{overpic}[width=1\columnwidth]{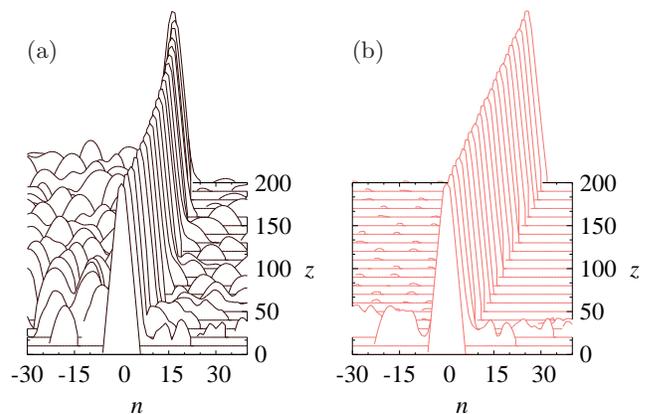}
\put(3,60){(a)}
\put(53,60){(b)}
\end{overpic}
\caption{(Color online) Amplitude profiles, on the logarithmic scale, of two
moving solitons, both kicked by $k=0.1$. Left: $P=15.93$; right: $P=16.36$.} 
\label{fig14}
\end{figure}

The difference in the mobility between the lighter and heavier solitons, kicked
with the same strength, is shown in more detail in Fig.~\ref{fig14}, where the
soliton field is displayed on the logarithmic scale: the shape distortion,
manifested by the generation of radiation tails, is greater for lighter
solitons [Fig.~\ref{fig14}(a)]. Thus, the heavier ones are more stable in the
traveling state [Fig.~\ref{fig14}(b)], as shown by the bottom inset in
Fig.~\ref{fig13}, which displays the persistent motion of the discrete soliton
with total power $P=16.4$.  Therefore, the heavier solitons are more
appropriate for applications related to transport properties.

\section{Collisions between moving solitons}

Once the regions with mobility of the discrete solitons having been
established, it is natural to consider collisions between two traveling robust
solitons, which are set in motion by kicks in opposite direction. The
corresponding input can be written as
\begin{equation}
\psi _{n}(z=0)=\phi _{n}^{(1)}e^{ik_{1}(n-n1)}+\phi
_{n}^{(2)}e^{ik_{2}(n-n2)},
\label{collide}
\end{equation}
where $n_{1}$ and $n_{2}$ are the initial positions of centers of the
stationary solutions, $\phi^{(1)}$ and $\phi^{(2)}$, respectively.

\begin{figure}[tb]
\begin{overpic}[width=1 \columnwidth]{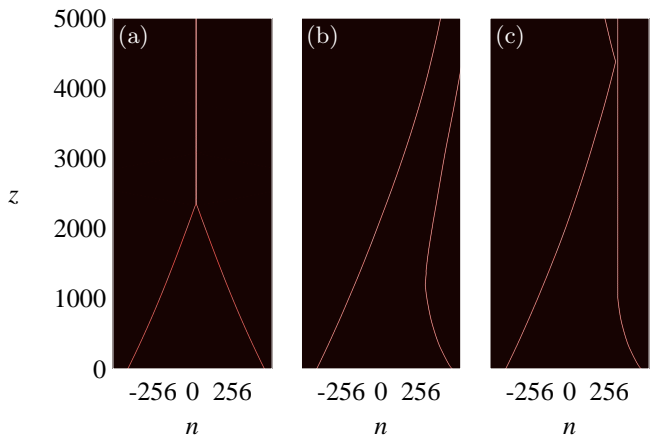}
\put(17,64){\white{(a)}}
\put(46,64){\white{(b)}}
\put(75,64){\white{(c)}}
\end{overpic}
\caption{(Color online) Three different scenarios of collisions between
traveling discrete solitons: merger for $k_{1}=-k_{2}=0.1$ (a), rebound for
$k_{1}=0.1$ and $k_{2}=-1.9$ (b), and the formation of the signal-blocker pair
for $k_{1}=0.1$ and $k_{2}=-0.2$ (c).}
\label{fig15}
\end{figure}

Figure~\ref{fig15} displays different scenarios of the interaction between
identical colliding modes. The first case [Fig.~\ref{fig15}(a)] shows the
interaction between two discrete solitons initially kicked with
$k_{1}=-k_{2}=0.1$.  After the collision, they merge into a single quiescent
structure with the double total power. In another case [Fig.~\ref{fig15}(b)],
the collision does not take place, due to repulsion between the solitons
(\textit{rebound}). The stronger kicked mode, with $k_{2}=-1.9$, changes the
direction of its velocity, and keeps moving almost at the same velocity as the
other mode, which was kicked by $k_{1}=0.1$. Finally, when the two modes are
kicked by $k_{1}=0.1$ and $k_{2}=-0.2$, the interaction between them looks as a
\textit{signal-blocker pair}, i.e., one discrete soliton gets trapped,
producing a barrier for the perfect reflection of the other soliton [see
Fig.~\ref{fig15}(c)].

The interaction picture observed in panels (b) and (c) of Fig.~\ref{fig15}
suggests that the interaction between the discrete solitons may feature
effectively long-range interactions. This may be explained by the interaction
of the phonon (radiation) tail of each soliton, such those observed in Fig.
\ref{fig14}, with the body of the other, cf. a similar mechanism considered
earlier in continuous media~\cite{radiation}.

\section{Conclusion}

In this work, we used the 1D DNLS model with the CQ (cubic-quintic) on-site
nonlinearity to investigate fundamental mobility regimes for discrete solitons.
The cubic term was taken, as usual, with the self-focusing sign, while both
signs of the quintic term were considered separately, as well as the usual
cubic model, when the quintic term is absent. The analytical part of the work
was based on the VA (variational approximation), with the aim to estimate the
mobility threshold, i.e., the largest value of the discrete-soliton's total
power (norm) which admits free motion of the kicked soliton. This was done by
comparing the largest possible value of the soliton's effective kinetic energy
with the height of the PN potential barrier. It is relevant to stress that this
analytical prediction is novel also for the usual DNLS equation with the purely
cubic nonlinearity. The prediction was found to be in a reasonable agreement
with numerical results. In addition, this analytical estimate qualitatively
predicts that the increase of the coefficient in front of self-focusing quintic
term quickly suppresses the mobility. In the DNLS model with competing
self-focusing and defocusing CQ terms, we have identified, by means of the
numerical analysis, regions in the parameter space where even and odd
fundamental modes exchange their stability, which involves the appearance of
intermediate asymmetric modes. In this case, stable solitons are kicked to
overcome the PN barrier and get in a state of persistent motion. On the other
hand, it has been demonstrate by means of direct simulations that unstable
solitons start progressive motion spontaneously, provided that they initially
exist above the mobility threshold. Finally, collisions between stable moving
discrete solitons were considered, and three different scenarios identified for
them.

The analysis reported here may be extended in other directions. In particular,
it may be interesting to consider possible mobility of excited
(non-fundamental) localized states, such as twisted modes~\cite{twisted}. A
challenging problem is to generalize the analysis for those two-dimensional
models where mobile modes may
exist~\cite{PhysRevE.73.046602},~\cite{PhysRevLett.99.214103}.

\section{Acknowledgment}

C. M.-C. thanks Uta Naether and Mario I. Molina for valuable discussions. We
acknowledge funding support from FONDECYT Grant 1110142, Programa ICM
P10-030-F, and Programa de Financiamiento Basal de CONICYT (FB0824/2008).
B.A.M. appreciates hospitality of the Faculty of Engineering and Applied
Sciences at the Universidad de los Andes (Santiago, Chile).


\begin{thebibliography}{57}
\expandafter\ifx\csname natexlab\endcsname\relax\def\natexlab#1{#1}\fi
\expandafter\ifx\csname bibnamefont\endcsname\relax
  \def\bibnamefont#1{#1}\fi
\expandafter\ifx\csname bibfnamefont\endcsname\relax
  \def\bibfnamefont#1{#1}\fi
\expandafter\ifx\csname citenamefont\endcsname\relax
  \def\citenamefont#1{#1}\fi
\expandafter\ifx\csname url\endcsname\relax
  \def\url#1{\texttt{#1}}\fi
\expandafter\ifx\csname urlprefix\endcsname\relax\def\urlprefix{URL }\fi
\providecommand{\bibinfo}[2]{#2}
\providecommand{\eprint}[2][]{\url{#2}}

\bibitem[{\citenamefont{Hasegawa}(2002)}]{Hasegawa2002}
\bibinfo{author}{\bibfnamefont{A.}~\bibnamefont{Hasegawa}},
  \bibinfo{journal}{Optics and Photonics News} \textbf{\bibinfo{volume}{13}},
  \bibinfo{pages}{33} (\bibinfo{year}{2002}), \bibinfo{issn}{1047-6938}.

\bibitem[{\citenamefont{Aitchison et~al.}(1991)\citenamefont{Aitchison,
  Silberberg, Weiner, Leaird, Oliver, Jackel, Vogel, and Smith}}]{Aitchison:91}
\bibinfo{author}{\bibfnamefont{J.~S.} \bibnamefont{Aitchison}},
  \bibinfo{author}{\bibfnamefont{Y.}~\bibnamefont{Silberberg}},
  \bibinfo{author}{\bibfnamefont{A.~M.} \bibnamefont{Weiner}},
  \bibinfo{author}{\bibfnamefont{D.~E.} \bibnamefont{Leaird}},
  \bibinfo{author}{\bibfnamefont{M.~K.} \bibnamefont{Oliver}},
  \bibinfo{author}{\bibfnamefont{J.~L.} \bibnamefont{Jackel}},
  \bibinfo{author}{\bibfnamefont{E.~M.} \bibnamefont{Vogel}}, \bibnamefont{and}
  \bibinfo{author}{\bibfnamefont{P.~W.~E.} \bibnamefont{Smith}},
  \bibinfo{journal}{J. Opt. Soc. Am. B} \textbf{\bibinfo{volume}{8}},
  \bibinfo{pages}{1290} (\bibinfo{year}{1991}).

\bibitem[{\citenamefont{Fleischer et~al.}(2003)\citenamefont{Fleischer, Segev,
  Efremidis, and Christodoulides}}]{fleis}
\bibinfo{author}{\bibfnamefont{J.~W.} \bibnamefont{Fleischer}},
  \bibinfo{author}{\bibfnamefont{M.}~\bibnamefont{Segev}},
  \bibinfo{author}{\bibfnamefont{N.~K.} \bibnamefont{Efremidis}},
  \bibnamefont{and} \bibinfo{author}{\bibfnamefont{D.~N.}
  \bibnamefont{Christodoulides}}, \bibinfo{journal}{Nature}
  \textbf{\bibinfo{volume}{422}}, \bibinfo{pages}{147} (\bibinfo{year}{2003}).

\bibitem[{\citenamefont{Christodoulides
  et~al.}(2003)\citenamefont{Christodoulides, Lederer, and
  Silberberg}}]{Christodoulides2003}
\bibinfo{author}{\bibfnamefont{D.~N.} \bibnamefont{Christodoulides}},
  \bibinfo{author}{\bibfnamefont{F.}~\bibnamefont{Lederer}}, \bibnamefont{and}
  \bibinfo{author}{\bibfnamefont{Y.}~\bibnamefont{Silberberg}},
  \bibinfo{journal}{Nature} \textbf{\bibinfo{volume}{424}},
  \bibinfo{pages}{817} (\bibinfo{year}{2003}), \bibinfo{issn}{1476-4687}.

\bibitem[{\citenamefont{Flach and Willis}(1998)}]{Flach1998181}
\bibinfo{author}{\bibfnamefont{S.}~\bibnamefont{Flach}} \bibnamefont{and}
  \bibinfo{author}{\bibfnamefont{C.~R.} \bibnamefont{Willis}},
  \bibinfo{journal}{Physics Reports} \textbf{\bibinfo{volume}{295}},
  \bibinfo{pages}{181 } (\bibinfo{year}{1998}), \bibinfo{issn}{0370-1573}.

\bibitem[{\citenamefont{Kevrekidis}(2009)}]{kevrekidis2009discrete}
\bibinfo{author}{\bibfnamefont{P.~G.} \bibnamefont{Kevrekidis}},
  \emph{\bibinfo{title}{The Discrete Nonlinear Schr{\"o}dinger Equation:
  Mathematical Analysis, Numerical Computations and Physical Perspectives}},
  Springer tracts in modern physics (\bibinfo{publisher}{Springer-Verlag Berlin
  Heidelberg}, \bibinfo{year}{2009}), ISBN \bibinfo{isbn}{9783540891994}.

\bibitem[{\citenamefont{Eisenberg et~al.}(2002)\citenamefont{Eisenberg,
  Morandotti, Silberberg, Arnold, Pennelli, and Aitchison}}]{Eisenberg:02}
\bibinfo{author}{\bibfnamefont{H.~S.} \bibnamefont{Eisenberg}},
  \bibinfo{author}{\bibfnamefont{R.}~\bibnamefont{Morandotti}},
  \bibinfo{author}{\bibfnamefont{Y.}~\bibnamefont{Silberberg}},
  \bibinfo{author}{\bibfnamefont{J.~M.} \bibnamefont{Arnold}},
  \bibinfo{author}{\bibfnamefont{G.}~\bibnamefont{Pennelli}}, \bibnamefont{and}
  \bibinfo{author}{\bibfnamefont{J.~S.} \bibnamefont{Aitchison}},
  \bibinfo{journal}{J. Opt. Soc. Am. B} \textbf{\bibinfo{volume}{19}},
  \bibinfo{pages}{2938} (\bibinfo{year}{2002}).

\bibitem[{\citenamefont{Efremidis et~al.}(2002)\citenamefont{Efremidis, Sears,
  Christodoulides, Fleischer, and Segev}}]{PhysRevE.66.046602}
\bibinfo{author}{\bibfnamefont{N.~K.} \bibnamefont{Efremidis}},
  \bibinfo{author}{\bibfnamefont{S.}~\bibnamefont{Sears}},
  \bibinfo{author}{\bibfnamefont{D.~N.} \bibnamefont{Christodoulides}},
  \bibinfo{author}{\bibfnamefont{J.~W.} \bibnamefont{Fleischer}},
  \bibnamefont{and} \bibinfo{author}{\bibfnamefont{M.}~\bibnamefont{Segev}},
  \bibinfo{journal}{Phys. Rev. E} \textbf{\bibinfo{volume}{66}},
  \bibinfo{pages}{046602} (\bibinfo{year}{2002}).

\bibitem[{\citenamefont{Sukhorukov et~al.}(2003)\citenamefont{Sukhorukov,
  Kivshar, Eisenberg, and Silberberg}}]{1158804}
\bibinfo{author}{\bibfnamefont{A.~A.} \bibnamefont{Sukhorukov}},
  \bibinfo{author}{\bibfnamefont{Y.~S.} \bibnamefont{Kivshar}},
  \bibinfo{author}{\bibfnamefont{H.~S.} \bibnamefont{Eisenberg}},
  \bibnamefont{and}
  \bibinfo{author}{\bibfnamefont{Y.}~\bibnamefont{Silberberg}},
  \bibinfo{journal}{Quantum Electronics, IEEE Journal of}
  \textbf{\bibinfo{volume}{39}}, \bibinfo{pages}{31 } (\bibinfo{year}{2003}),
  \bibinfo{issn}{0018-9197}.

\bibitem[{\citenamefont{Cataliotti et~al.}(2003)\citenamefont{Cataliotti,
  Fallani, Ferlaino, Fort, Maddaloni, and Inguscio}}]{1367-2630-5-1-371}
\bibinfo{author}{\bibfnamefont{F.~S.} \bibnamefont{Cataliotti}},
  \bibinfo{author}{\bibfnamefont{L.}~\bibnamefont{Fallani}},
  \bibinfo{author}{\bibfnamefont{F.}~\bibnamefont{Ferlaino}},
  \bibinfo{author}{\bibfnamefont{C.}~\bibnamefont{Fort}},
  \bibinfo{author}{\bibfnamefont{P.}~\bibnamefont{Maddaloni}},
  \bibnamefont{and} \bibinfo{author}{\bibfnamefont{M.}~\bibnamefont{Inguscio}},
  \bibinfo{journal}{New Journal of Physics} \textbf{\bibinfo{volume}{5}},
  \bibinfo{pages}{71} (\bibinfo{year}{2003}).

\bibitem[{\citenamefont{Abdullaev et~al.}(2001)\citenamefont{Abdullaev,
  Baizakov, Darmanyan, Konotop, and Salerno}}]{PhysRevA.64.043606}
\bibinfo{author}{\bibfnamefont{F.~K.} \bibnamefont{Abdullaev}},
  \bibinfo{author}{\bibfnamefont{B.~B.} \bibnamefont{Baizakov}},
  \bibinfo{author}{\bibfnamefont{S.~A.} \bibnamefont{Darmanyan}},
  \bibinfo{author}{\bibfnamefont{V.~V.} \bibnamefont{Konotop}},
  \bibnamefont{and} \bibinfo{author}{\bibfnamefont{M.}~\bibnamefont{Salerno}},
  \bibinfo{journal}{Phys. Rev. A} \textbf{\bibinfo{volume}{64}},
  \bibinfo{pages}{043606} (\bibinfo{year}{2001}).

\bibitem[{\citenamefont{Alfimov et~al.}(2002)\citenamefont{Alfimov, Kevrekidis,
  Konotop, and Salerno}}]{PhysRevE.66.046608}
\bibinfo{author}{\bibfnamefont{G.~L.} \bibnamefont{Alfimov}},
  \bibinfo{author}{\bibfnamefont{P.~G.} \bibnamefont{Kevrekidis}},
  \bibinfo{author}{\bibfnamefont{V.~V.} \bibnamefont{Konotop}},
  \bibnamefont{and} \bibinfo{author}{\bibfnamefont{M.}~\bibnamefont{Salerno}},
  \bibinfo{journal}{Phys. Rev. E} \textbf{\bibinfo{volume}{66}},
  \bibinfo{pages}{046608} (\bibinfo{year}{2002}).

\bibitem[{\citenamefont{Gligori\'{c} et~al.}(2008)\citenamefont{Gligori\'{c},
  Maluckov, Had\v{z}ievski, and Malomed}}]{Sandra1D}
\bibinfo{author}{\bibfnamefont{G.}~\bibnamefont{Gligori\'{c}}},
  \bibinfo{author}{\bibfnamefont{A.}~\bibnamefont{Maluckov}},
  \bibinfo{author}{\bibfnamefont{L.}~\bibnamefont{Had\v{z}ievski}},
  \bibnamefont{and} \bibinfo{author}{\bibfnamefont{B.~A.}
  \bibnamefont{Malomed}}, \bibinfo{journal}{Phys. Rev. A}
  \textbf{\bibinfo{volume}{78}}, \bibinfo{pages}{063615}
  (\bibinfo{year}{2008}).

\bibitem[{\citenamefont{Gligori\'{c} et~al.}(2010)\citenamefont{Gligori\'{c},
  Maluckov, Had\v{z}ievski, and Malomed}}]{Sandra2D}
\bibinfo{author}{\bibfnamefont{G.}~\bibnamefont{Gligori\'{c}}},
  \bibinfo{author}{\bibfnamefont{A.}~\bibnamefont{Maluckov}},
  \bibinfo{author}{\bibfnamefont{L.}~\bibnamefont{Had\v{z}ievski}},
  \bibnamefont{and} \bibinfo{author}{\bibfnamefont{B.~A.}
  \bibnamefont{Malomed}}, \bibinfo{journal}{Phys. Rev. A}
  \textbf{\bibinfo{volume}{81}}, \bibinfo{pages}{013633}
  (\bibinfo{year}{2010}).

\bibitem[{\citenamefont{Ustinov et~al.}(1993)\citenamefont{Ustinov, Cirillo,
  and Malomed}}]{Matteo}
\bibinfo{author}{\bibfnamefont{A.~V.} \bibnamefont{Ustinov}},
  \bibinfo{author}{\bibfnamefont{M.}~\bibnamefont{Cirillo}}, \bibnamefont{and}
  \bibinfo{author}{\bibfnamefont{B.~A.} \bibnamefont{Malomed}},
  \bibinfo{journal}{Phys. Rev. B} \textbf{\bibinfo{volume}{47}},
  \bibinfo{pages}{8357} (\bibinfo{year}{1993}).

\bibitem[{\citenamefont{Fistul}(2003)}]{fistul:725}
\bibinfo{author}{\bibfnamefont{M.~V.} \bibnamefont{Fistul}},
  \bibinfo{journal}{Chaos: An Interdisciplinary Journal of Nonlinear Science}
  \textbf{\bibinfo{volume}{13}}, \bibinfo{pages}{725} (\bibinfo{year}{2003}).

\bibitem[{\citenamefont{Mazo and Orlando}(2003)}]{mazo:733}
\bibinfo{author}{\bibfnamefont{J.~J.} \bibnamefont{Mazo}} \bibnamefont{and}
  \bibinfo{author}{\bibfnamefont{T.~P.} \bibnamefont{Orlando}},
  \bibinfo{journal}{Chaos: An Interdisciplinary Journal of Nonlinear Science}
  \textbf{\bibinfo{volume}{13}}, \bibinfo{pages}{733} (\bibinfo{year}{2003}).

\bibitem[{\citenamefont{Dauxois et~al.}(1993)\citenamefont{Dauxois, Peyrard,
  and Bishop}}]{PhysRevE.47.R44}
\bibinfo{author}{\bibfnamefont{T.}~\bibnamefont{Dauxois}},
  \bibinfo{author}{\bibfnamefont{M.}~\bibnamefont{Peyrard}}, \bibnamefont{and}
  \bibinfo{author}{\bibfnamefont{A.~R.} \bibnamefont{Bishop}},
  \bibinfo{journal}{Phys. Rev. E} \textbf{\bibinfo{volume}{47}},
  \bibinfo{pages}{R44} (\bibinfo{year}{1993}).

\bibitem[{\citenamefont{Peyrard et~al.}(1993)\citenamefont{Peyrard, Dauxois,
  Hoyet, and Willis}}]{Peyrard1993104}
\bibinfo{author}{\bibfnamefont{M.}~\bibnamefont{Peyrard}},
  \bibinfo{author}{\bibfnamefont{T.}~\bibnamefont{Dauxois}},
  \bibinfo{author}{\bibfnamefont{H.}~\bibnamefont{Hoyet}}, \bibnamefont{and}
  \bibinfo{author}{\bibfnamefont{C.}~\bibnamefont{Willis}},
  \bibinfo{journal}{Physica D}
  \textbf{\bibinfo{volume}{68}}, \bibinfo{pages}{104 } (\bibinfo{year}{1993}),
  \bibinfo{issn}{0167-2789}.

\bibitem[{\citenamefont{Aceves et~al.}(1994)\citenamefont{Aceves, Angelis,
  Trillo, and Wabnitz}}]{Aceves:94}
\bibinfo{author}{\bibfnamefont{A.~B.} \bibnamefont{Aceves}},
  \bibinfo{author}{\bibfnamefont{C.~D.} \bibnamefont{Angelis}},
  \bibinfo{author}{\bibfnamefont{S.}~\bibnamefont{Trillo}}, \bibnamefont{and}
  \bibinfo{author}{\bibfnamefont{S.}~\bibnamefont{Wabnitz}},
  \bibinfo{journal}{Opt. Lett.} \textbf{\bibinfo{volume}{19}},
  \bibinfo{pages}{332} (\bibinfo{year}{1994}).

\bibitem[{\citenamefont{Vicencio et~al.}(2003)\citenamefont{Vicencio, Molina,
  and Kivshar}}]{Vicencio:03}
\bibinfo{author}{\bibfnamefont{R.~A.} \bibnamefont{Vicencio}},
  \bibinfo{author}{\bibfnamefont{M.~I.} \bibnamefont{Molina}},
  \bibnamefont{and} \bibinfo{author}{\bibfnamefont{Y.~S.}
  \bibnamefont{Kivshar}}, \bibinfo{journal}{Opt. Lett.}
  \textbf{\bibinfo{volume}{28}}, \bibinfo{pages}{1942} (\bibinfo{year}{2003}).

\bibitem[{\citenamefont{Cuevas et~al.}(2005)\citenamefont{Cuevas, Malomed, ,
  and Kevrekidis}}]{Cuevas}
\bibinfo{author}{\bibfnamefont{J.}~\bibnamefont{Cuevas}},
  \bibinfo{author}{\bibfnamefont{B.~A.} \bibnamefont{Malomed}}, ,
  \bibnamefont{and} \bibinfo{author}{\bibfnamefont{P.~G.}
  \bibnamefont{Kevrekidis}}, \bibinfo{journal}{Phys. Rev. E}
  \textbf{\bibinfo{volume}{71}}, \bibinfo{pages}{066614}
  (\bibinfo{year}{2005}).

\bibitem[{\citenamefont{Morandotti et~al.}(1999)\citenamefont{Morandotti,
  Peschel, Aitchison, Eisenberg, and Silberberg}}]{PhysRevLett.83.2726}
\bibinfo{author}{\bibfnamefont{R.}~\bibnamefont{Morandotti}},
  \bibinfo{author}{\bibfnamefont{U.}~\bibnamefont{Peschel}},
  \bibinfo{author}{\bibfnamefont{J.~S.} \bibnamefont{Aitchison}},
  \bibinfo{author}{\bibfnamefont{H.~S.} \bibnamefont{Eisenberg}},
  \bibnamefont{and}
  \bibinfo{author}{\bibfnamefont{Y.}~\bibnamefont{Silberberg}},
  \bibinfo{journal}{Phys. Rev. Lett.} \textbf{\bibinfo{volume}{83}},
  \bibinfo{pages}{2726} (\bibinfo{year}{1999}).

\bibitem[{\citenamefont{Peschel et~al.}(2002)\citenamefont{Peschel, Morandotti,
  Arnold, Aitchison, Eisenberg, Silberberg, Pertsch, and Lederer}}]{Peschel:02}
\bibinfo{author}{\bibfnamefont{U.}~\bibnamefont{Peschel}},
  \bibinfo{author}{\bibfnamefont{R.}~\bibnamefont{Morandotti}},
  \bibinfo{author}{\bibfnamefont{J.~M.} \bibnamefont{Arnold}},
  \bibinfo{author}{\bibfnamefont{J.~S.} \bibnamefont{Aitchison}},
  \bibinfo{author}{\bibfnamefont{H.~S.} \bibnamefont{Eisenberg}},
  \bibinfo{author}{\bibfnamefont{Y.}~\bibnamefont{Silberberg}},
  \bibinfo{author}{\bibfnamefont{T.}~\bibnamefont{Pertsch}}, \bibnamefont{and}
  \bibinfo{author}{\bibfnamefont{F.}~\bibnamefont{Lederer}},
  \bibinfo{journal}{J. Opt. Soc. Am. B} \textbf{\bibinfo{volume}{19}},
  \bibinfo{pages}{2637} (\bibinfo{year}{2002}).

\bibitem[{\citenamefont{Vicencio and Johansson}(2006)}]{PhysRevE.73.046602}
\bibinfo{author}{\bibfnamefont{R.~A.} \bibnamefont{Vicencio}} \bibnamefont{and}
  \bibinfo{author}{\bibfnamefont{M.}~\bibnamefont{Johansson}},
  \bibinfo{journal}{Phys. Rev. E} \textbf{\bibinfo{volume}{73}},
  \bibinfo{pages}{046602} (\bibinfo{year}{2006}).

\bibitem[{\citenamefont{Susanto et~al.}(2007)\citenamefont{Susanto, Kevrekidis,
  Carretero-GonzÃ¡lez, Malomed, and Frantzeskakis}}]{PhysRevLett.99.214103}
\bibinfo{author}{\bibfnamefont{H.}~\bibnamefont{Susanto}},
  \bibinfo{author}{\bibfnamefont{P.~G.} \bibnamefont{Kevrekidis}},
  \bibinfo{author}{\bibfnamefont{R.}~\bibnamefont{Carretero-GonzÃ¡lez}},
  \bibinfo{author}{\bibfnamefont{B.~A.} \bibnamefont{Malomed}},
  \bibnamefont{and} \bibinfo{author}{\bibfnamefont{D.~J.}
  \bibnamefont{Frantzeskakis}}, \bibinfo{journal}{Phys. Rev. Lett.}
  \textbf{\bibinfo{volume}{99}}, \bibinfo{pages}{214103}
  (\bibinfo{year}{2007}).

\bibitem[{\citenamefont{Naether
  et~al.}(2011{\natexlab{a}})\citenamefont{Naether, Vicencio, and
  Stepi\'{c}}}]{Naether:11}
\bibinfo{author}{\bibfnamefont{U.}~\bibnamefont{Naether}},
  \bibinfo{author}{\bibfnamefont{R.~A.} \bibnamefont{Vicencio}},
  \bibnamefont{and}
  \bibinfo{author}{\bibfnamefont{M.}~\bibnamefont{Stepi\'{c}}},
  \bibinfo{journal}{Opt. Lett.} \textbf{\bibinfo{volume}{36}},
  \bibinfo{pages}{1467} (\bibinfo{year}{2011}{\natexlab{a}}).

\bibitem[{\citenamefont{Egorov and Lederer}(2013)}]{Egorov:13}
\bibinfo{author}{\bibfnamefont{O.~A.} \bibnamefont{Egorov}} \bibnamefont{and}
  \bibinfo{author}{\bibfnamefont{F.}~\bibnamefont{Lederer}},
  \bibinfo{journal}{Opt. Lett.} \textbf{\bibinfo{volume}{38}},
  \bibinfo{pages}{1010} (\bibinfo{year}{2013}).

\bibitem[{\citenamefont{Maluckov et~al.}(2008)\citenamefont{Maluckov,
  Had\v{z}ievski, and Malomed}}]{PhysRevE.77.036604}
\bibinfo{author}{\bibfnamefont{A.}~\bibnamefont{Maluckov}},
  \bibinfo{author}{\bibfnamefont{L.}~\bibnamefont{Had\v{z}ievski}},
  \bibnamefont{and} \bibinfo{author}{\bibfnamefont{B.~A.}
  \bibnamefont{Malomed}}, \bibinfo{journal}{Phys. Rev. E}
  \textbf{\bibinfo{volume}{77}}, \bibinfo{pages}{036604}
  (\bibinfo{year}{2008}).

\bibitem[{\citenamefont{Campbell et~al.}(2004)\citenamefont{Campbell, Flach,
  and Kivshar}}]{campbell:43}
\bibinfo{author}{\bibfnamefont{D.~K.} \bibnamefont{Campbell}},
  \bibinfo{author}{\bibfnamefont{S.}~\bibnamefont{Flach}}, \bibnamefont{and}
  \bibinfo{author}{\bibfnamefont{Y.~S.} \bibnamefont{Kivshar}},
  \bibinfo{journal}{Physics Today} \textbf{\bibinfo{volume}{57}},
  \bibinfo{pages}{43} (\bibinfo{year}{2004}).

\bibitem[{\citenamefont{Lederer et~al.}(2008)\citenamefont{Lederer, Stegeman,
  Christodoulides, Assanto, Segev, and Silberberg}}]{Lederer20081}
\bibinfo{author}{\bibfnamefont{F.}~\bibnamefont{Lederer}},
  \bibinfo{author}{\bibfnamefont{G.~I.} \bibnamefont{Stegeman}},
  \bibinfo{author}{\bibfnamefont{D.~N.} \bibnamefont{Christodoulides}},
  \bibinfo{author}{\bibfnamefont{G.}~\bibnamefont{Assanto}},
  \bibinfo{author}{\bibfnamefont{M.}~\bibnamefont{Segev}}, \bibnamefont{and}
  \bibinfo{author}{\bibfnamefont{Y.}~\bibnamefont{Silberberg}},
  \bibinfo{journal}{Physics Reports} \textbf{\bibinfo{volume}{463}},
  \bibinfo{pages}{1 } (\bibinfo{year}{2008}).

\bibitem[{\citenamefont{Flach and Gorbach}(2008)}]{Flach20081}
\bibinfo{author}{\bibfnamefont{S.}~\bibnamefont{Flach}} \bibnamefont{and}
  \bibinfo{author}{\bibfnamefont{A.~V.} \bibnamefont{Gorbach}},
  \bibinfo{journal}{Physics Reports} \textbf{\bibinfo{volume}{467}},
  \bibinfo{pages}{1 } (\bibinfo{year}{2008}).

\bibitem[{\citenamefont{Elstner et~al.}(1998)\citenamefont{Elstner, Porezag,
  Jungnickel, Elsner, Haugk, Frauenheim, Suhai, and
  Seifert}}]{PhysRevB.58.7260}
\bibinfo{author}{\bibfnamefont{M.}~\bibnamefont{Elstner}},
  \bibinfo{author}{\bibfnamefont{D.}~\bibnamefont{Porezag}},
  \bibinfo{author}{\bibfnamefont{G.}~\bibnamefont{Jungnickel}},
  \bibinfo{author}{\bibfnamefont{J.}~\bibnamefont{Elsner}},
  \bibinfo{author}{\bibfnamefont{M.}~\bibnamefont{Haugk}},
  \bibinfo{author}{\bibfnamefont{T.}~\bibnamefont{Frauenheim}},
  \bibinfo{author}{\bibfnamefont{S.}~\bibnamefont{Suhai}}, \bibnamefont{and}
  \bibinfo{author}{\bibfnamefont{G.}~\bibnamefont{Seifert}},
  \bibinfo{journal}{Phys. Rev. B} \textbf{\bibinfo{volume}{58}},
  \bibinfo{pages}{7260} (\bibinfo{year}{1998}).

\bibitem[{\citenamefont{Carretero-GonzÃ¡lez
  et~al.}(2006)\citenamefont{Carretero-GonzÃ¡lez, Talley, Chong, and
  Malomed}}]{CarreteroGonzalez200677}
\bibinfo{author}{\bibfnamefont{R.}~\bibnamefont{Carretero-GonzÃ¡lez}},
  \bibinfo{author}{\bibfnamefont{J.~D.} \bibnamefont{Talley}},
  \bibinfo{author}{\bibfnamefont{C.}~\bibnamefont{Chong}}, \bibnamefont{and}
  \bibinfo{author}{\bibfnamefont{B.~A.} \bibnamefont{Malomed}},
  \bibinfo{journal}{Physica D}
  \textbf{\bibinfo{volume}{216}}, \bibinfo{pages}{77 } (\bibinfo{year}{2006}),
  \bibinfo{issn}{0167-2789}.

\bibitem[{\citenamefont{Chong et~al.}(2009)\citenamefont{Chong,
  Carretero-GonzÃ¡lez, Malomed, and Kevrekidis}}]{Chong2009126}
\bibinfo{author}{\bibfnamefont{C.}~\bibnamefont{Chong}},
  \bibinfo{author}{\bibfnamefont{R.}~\bibnamefont{Carretero-GonzÃ¡lez}},
  \bibinfo{author}{\bibfnamefont{B.~A.} \bibnamefont{Malomed}},
  \bibnamefont{and} \bibinfo{author}{\bibfnamefont{P.~G.}
  \bibnamefont{Kevrekidis}}, \bibinfo{journal}{Physica D}
  \textbf{\bibinfo{volume}{238}}, \bibinfo{pages}{126 } (\bibinfo{year}{2009}).

\bibitem[{\citenamefont{Lederer et~al.}(2001)\citenamefont{Lederer, Darmanyan,
  and Kobyakov}}]{Falk_spring}
\bibinfo{author}{\bibfnamefont{F.}~\bibnamefont{Lederer}},
  \bibinfo{author}{\bibfnamefont{S.}~\bibnamefont{Darmanyan}},
  \bibnamefont{and} \bibinfo{author}{\bibfnamefont{A.}~\bibnamefont{Kobyakov}},
  in \emph{\bibinfo{booktitle}{Nonlinearity and Disorder: Theory and
  Applications}}, edited by
  \bibinfo{editor}{\bibfnamefont{F.}~\bibnamefont{Abdullaev}},
  \bibinfo{editor}{\bibfnamefont{O.}~\bibnamefont{Bang}}, \bibnamefont{and}
  \bibinfo{editor}{\bibfnamefont{M.}~\bibnamefont{S\"orensen}}
  (\bibinfo{publisher}{Springer Netherlands}, \bibinfo{year}{2001}),
  vol.~\bibinfo{volume}{45} of \emph{\bibinfo{series}{NATO Science Series}},
  pp. \bibinfo{pages}{131--157}, ISBN \bibinfo{isbn}{978-1-4020-0192-5}.

\bibitem[{\citenamefont{Khare et~al.}(2005)\citenamefont{Khare, Rasmussen,
  Samuelsen, and Saxena}}]{0305-4470-38-4-002}
\bibinfo{author}{\bibfnamefont{A.}~\bibnamefont{Khare}},
  \bibinfo{author}{\bibfnamefont{K.~Ã.} \bibnamefont{Rasmussen}},
  \bibinfo{author}{\bibfnamefont{M.~R.} \bibnamefont{Samuelsen}},
  \bibnamefont{and} \bibinfo{author}{\bibfnamefont{A.}~\bibnamefont{Saxena}},
  \bibinfo{journal}{Journal of Physics A: Mathematical and General}
  \textbf{\bibinfo{volume}{38}}, \bibinfo{pages}{807} (\bibinfo{year}{2005}).

\bibitem[{\citenamefont{Kivshar and Campbell}(1993)}]{PhysRevE.48.3077}
\bibinfo{author}{\bibfnamefont{Y.~S.} \bibnamefont{Kivshar}} \bibnamefont{and}
  \bibinfo{author}{\bibfnamefont{D.~K.} \bibnamefont{Campbell}},
  \bibinfo{journal}{Phys. Rev. E} \textbf{\bibinfo{volume}{48}},
  \bibinfo{pages}{3077} (\bibinfo{year}{1993}).

\bibitem[{\citenamefont{Guzm\'an-Silva
  et~al.}(2013)\citenamefont{Guzm\'an-Silva, Lou, Naether, R\"uter, Kip, and
  Vicencio}}]{PhysRevA.87.043837}
\bibinfo{author}{\bibfnamefont{D.}~\bibnamefont{Guzm\'an-Silva}},
  \bibinfo{author}{\bibfnamefont{C.}~\bibnamefont{Lou}},
  \bibinfo{author}{\bibfnamefont{U.}~\bibnamefont{Naether}},
  \bibinfo{author}{\bibfnamefont{C.~E.} \bibnamefont{R\"uter}},
  \bibinfo{author}{\bibfnamefont{D.}~\bibnamefont{Kip}}, \bibnamefont{and}
  \bibinfo{author}{\bibfnamefont{R.~A.} \bibnamefont{Vicencio}},
  \bibinfo{journal}{Phys. Rev. A} \textbf{\bibinfo{volume}{87}},
  \bibinfo{pages}{043837} (\bibinfo{year}{2013}).

\bibitem[{\citenamefont{Naether
  et~al.}(2011{\natexlab{b}})\citenamefont{Naether, Vicencio, and
  Johansson}}]{PhysRevE.83.036601}
\bibinfo{author}{\bibfnamefont{U.}~\bibnamefont{Naether}},
  \bibinfo{author}{\bibfnamefont{R.~A.} \bibnamefont{Vicencio}},
  \bibnamefont{and}
  \bibinfo{author}{\bibfnamefont{M.}~\bibnamefont{Johansson}},
  \bibinfo{journal}{Phys. Rev. E} \textbf{\bibinfo{volume}{83}},
  \bibinfo{pages}{036601} (\bibinfo{year}{2011}{\natexlab{b}}).

\bibitem[{\citenamefont{Rojas-Rojas et~al.}(2011)\citenamefont{Rojas-Rojas,
  Vicencio, Molina, and Abdullaev}}]{PhysRevA.84.033621}
\bibinfo{author}{\bibfnamefont{S.}~\bibnamefont{Rojas-Rojas}},
  \bibinfo{author}{\bibfnamefont{R.~A.} \bibnamefont{Vicencio}},
  \bibinfo{author}{\bibfnamefont{M.~I.} \bibnamefont{Molina}},
  \bibnamefont{and} \bibinfo{author}{\bibfnamefont{F.~K.}
  \bibnamefont{Abdullaev}}, \bibinfo{journal}{Phys. Rev. A}
  \textbf{\bibinfo{volume}{84}}, \bibinfo{pages}{033621}
  (\bibinfo{year}{2011}).

\bibitem[{\citenamefont{Vicencio and Johansson}(2013)}]{PhysRevA.87.061803}
\bibinfo{author}{\bibfnamefont{R.~A.} \bibnamefont{Vicencio}} \bibnamefont{and}
  \bibinfo{author}{\bibfnamefont{M.}~\bibnamefont{Johansson}},
  \bibinfo{journal}{Phys. Rev. A} \textbf{\bibinfo{volume}{87}},
  \bibinfo{pages}{061803} (\bibinfo{year}{2013}).

\bibitem[{\citenamefont{Duncan et~al.}(1993)\citenamefont{Duncan, Eilbeck,
  Feddersen, and Wattis}}]{Duncan19931}
\bibinfo{author}{\bibfnamefont{D.~B.} \bibnamefont{Duncan}},
  \bibinfo{author}{\bibfnamefont{J.~C.} \bibnamefont{Eilbeck}},
  \bibinfo{author}{\bibfnamefont{H.}~\bibnamefont{Feddersen}},
  \bibnamefont{and} \bibinfo{author}{\bibfnamefont{J.~A.~D.}
  \bibnamefont{Wattis}}, \bibinfo{journal}{Physica D}
  \textbf{\bibinfo{volume}{68}}, \bibinfo{pages}{1 } (\bibinfo{year}{1993}),
  \bibinfo{issn}{0167-2789}.

\bibitem[{\citenamefont{Flach and Kladko}(1999)}]{Flach199961}
\bibinfo{author}{\bibfnamefont{S.}~\bibnamefont{Flach}} \bibnamefont{and}
  \bibinfo{author}{\bibfnamefont{K.}~\bibnamefont{Kladko}},
  \bibinfo{journal}{Physica D}
  \textbf{\bibinfo{volume}{127}}, \bibinfo{pages}{61 } (\bibinfo{year}{1999}),
  \bibinfo{issn}{0167-2789}.

\bibitem[{\citenamefont{Flach et~al.}(1999)\citenamefont{Flach, Zolotaryuk, and
  Kladko}}]{PhysRevE.59.6105}
\bibinfo{author}{\bibfnamefont{S.}~\bibnamefont{Flach}},
  \bibinfo{author}{\bibfnamefont{Y.}~\bibnamefont{Zolotaryuk}},
  \bibnamefont{and} \bibinfo{author}{\bibfnamefont{K.}~\bibnamefont{Kladko}},
  \bibinfo{journal}{Phys. Rev. E} \textbf{\bibinfo{volume}{59}},
  \bibinfo{pages}{6105} (\bibinfo{year}{1999}).

\bibitem[{\citenamefont{Ablowitz et~al.}(2002)\citenamefont{Ablowitz,
  Musslimani, and Biondini}}]{PhysRevE.65.026602}
\bibinfo{author}{\bibfnamefont{M.~J.} \bibnamefont{Ablowitz}},
  \bibinfo{author}{\bibfnamefont{Z.~H.} \bibnamefont{Musslimani}},
  \bibnamefont{and} \bibinfo{author}{\bibfnamefont{G.}~\bibnamefont{Biondini}},
  \bibinfo{journal}{Phys. Rev. E} \textbf{\bibinfo{volume}{65}},
  \bibinfo{pages}{026602} (\bibinfo{year}{2002}).

\bibitem[{\citenamefont{Papacharalampous
  et~al.}(2003)\citenamefont{Papacharalampous, Kevrekidis, Malomed, and
  Frantzeskakis}}]{PhysRevE.68.046604}
\bibinfo{author}{\bibfnamefont{I.~E.} \bibnamefont{Papacharalampous}},
  \bibinfo{author}{\bibfnamefont{P.~G.} \bibnamefont{Kevrekidis}},
  \bibinfo{author}{\bibfnamefont{B.~A.} \bibnamefont{Malomed}},
  \bibnamefont{and} \bibinfo{author}{\bibfnamefont{D.~J.}
  \bibnamefont{Frantzeskakis}}, \bibinfo{journal}{Phys. Rev. E}
  \textbf{\bibinfo{volume}{68}}, \bibinfo{pages}{046604}
  (\bibinfo{year}{2003}).

\bibitem[{\citenamefont{Vakhitov and Kolokolov}(1973)}]{VKcrit}
\bibinfo{author}{\bibfnamefont{N.~G.} \bibnamefont{Vakhitov}} \bibnamefont{and}
  \bibinfo{author}{\bibfnamefont{A.~A.} \bibnamefont{Kolokolov}},
  \bibinfo{journal}{Radiophysics and Quantum Electronics}
  \textbf{\bibinfo{volume}{16}}, \bibinfo{pages}{783} (\bibinfo{year}{1973}),
  \bibinfo{issn}{0033-8443}.

\bibitem[{\citenamefont{Burke and Knobloch}(2007)}]{burke:037102}
\bibinfo{author}{\bibfnamefont{J.}~\bibnamefont{Burke}} \bibnamefont{and}
  \bibinfo{author}{\bibfnamefont{E.}~\bibnamefont{Knobloch}},
  \bibinfo{journal}{Chaos: An Interdisciplinary Journal of Nonlinear Science}
  \textbf{\bibinfo{volume}{17}}, \bibinfo{eid}{037102}
  (pages~\bibinfo{numpages}{15}) (\bibinfo{year}{2007}).

\bibitem[{\citenamefont{Herring et~al.}(2005)\citenamefont{Herring, Kevrekidis,
  Carretero-GonzÃ¡lez, Malomed, Frantzeskakis, and Bishop}}]{Herring2005144}
\bibinfo{author}{\bibfnamefont{G.}~\bibnamefont{Herring}},
  \bibinfo{author}{\bibfnamefont{P.~G.} \bibnamefont{Kevrekidis}},
  \bibinfo{author}{\bibfnamefont{R.}~\bibnamefont{Carretero-GonzÃ¡lez}},
  \bibinfo{author}{\bibfnamefont{B.~A.} \bibnamefont{Malomed}},
  \bibinfo{author}{\bibfnamefont{D.~J.} \bibnamefont{Frantzeskakis}},
  \bibnamefont{and} \bibinfo{author}{\bibfnamefont{A.~R.}
  \bibnamefont{Bishop}}, \bibinfo{journal}{Physics Letters A}
  \textbf{\bibinfo{volume}{345}}, \bibinfo{pages}{144 } (\bibinfo{year}{2005}),
  \bibinfo{issn}{0375-9601}.

\bibitem[{\citenamefont{Sacchetti}(2009)}]{PhysRevLett.103.194101}
\bibinfo{author}{\bibfnamefont{A.}~\bibnamefont{Sacchetti}},
  \bibinfo{journal}{Phys. Rev. Lett.} \textbf{\bibinfo{volume}{103}},
  \bibinfo{pages}{194101} (\bibinfo{year}{2009}).

\bibitem[{\citenamefont{Had\v{z}ievski
  et~al.}(2004)\citenamefont{Had\v{z}ievski, Maluckov, Stepi\'{c}, and
  Kip}}]{PhysRevLett.93.033901}
\bibinfo{author}{\bibfnamefont{L.}~\bibnamefont{Had\v{z}ievski}},
  \bibinfo{author}{\bibfnamefont{A.}~\bibnamefont{Maluckov}},
  \bibinfo{author}{\bibfnamefont{M.}~\bibnamefont{Stepi\'{c}}},
  \bibnamefont{and} \bibinfo{author}{\bibfnamefont{D.}~\bibnamefont{Kip}},
  \bibinfo{journal}{Phys. Rev. Lett.} \textbf{\bibinfo{volume}{93}},
  \bibinfo{pages}{033901} (\bibinfo{year}{2004}).

\bibitem[{\citenamefont{Molina et~al.}(2006)\citenamefont{Molina, Vicencio, and
  Kivshar}}]{Molina:06}
\bibinfo{author}{\bibfnamefont{M.~I.} \bibnamefont{Molina}},
  \bibinfo{author}{\bibfnamefont{R.~A.} \bibnamefont{Vicencio}},
  \bibnamefont{and} \bibinfo{author}{\bibfnamefont{Y.~S.}
  \bibnamefont{Kivshar}}, \bibinfo{journal}{Opt. Lett.}
  \textbf{\bibinfo{volume}{31}}, \bibinfo{pages}{1693} (\bibinfo{year}{2006}).

\bibitem[{\citenamefont{Rosberg et~al.}(2006)\citenamefont{Rosberg, Neshev,
  Kr\'{o}likowski, Mitchell, Vicencio, Molina, and
  Kivshar}}]{PhysRevLett.97.083901}
\bibinfo{author}{\bibfnamefont{C.~R.} \bibnamefont{Rosberg}},
  \bibinfo{author}{\bibfnamefont{D.~N.} \bibnamefont{Neshev}},
  \bibinfo{author}{\bibfnamefont{W.}~\bibnamefont{Kr\'{o}likowski}},
  \bibinfo{author}{\bibfnamefont{A.}~\bibnamefont{Mitchell}},
  \bibinfo{author}{\bibfnamefont{R.~A.} \bibnamefont{Vicencio}},
  \bibinfo{author}{\bibfnamefont{M.~I.} \bibnamefont{Molina}},
  \bibnamefont{and} \bibinfo{author}{\bibfnamefont{Y.~S.}
  \bibnamefont{Kivshar}}, \bibinfo{journal}{Phys. Rev. Lett.}
  \textbf{\bibinfo{volume}{97}}, \bibinfo{pages}{083901}
  (\bibinfo{year}{2006}).

\bibitem[{\citenamefont{Arnol'd}(1989)}]{arnol1989mathematical}
\bibinfo{author}{\bibfnamefont{V.~I.} \bibnamefont{Arnol'd}},
  \emph{\bibinfo{title}{Mathematical Methods of Classical Mechanics}}, Graduate
  Texts in Mathematics (\bibinfo{publisher}{Springer}, \bibinfo{year}{1989}),
  ISBN \bibinfo{isbn}{9780387968902}.

\bibitem[{\citenamefont{Malomed}(1993)}]{radiation}
\bibinfo{author}{\bibfnamefont{B.~A.} \bibnamefont{Malomed}},
  \bibinfo{journal}{Phys. Rev. E} \textbf{\bibinfo{volume}{47}},
  \bibinfo{pages}{2874} (\bibinfo{year}{1993}).

\bibitem[{\citenamefont{Darmanyan et~al.}(1998)\citenamefont{Darmanyan,
  Kobyakov, and Lederer}}]{twisted}
\bibinfo{author}{\bibfnamefont{S.}~\bibnamefont{Darmanyan}},
  \bibinfo{author}{\bibfnamefont{A.}~\bibnamefont{Kobyakov}}, \bibnamefont{and}
  \bibinfo{author}{\bibfnamefont{F.}~\bibnamefont{Lederer}},
  \bibinfo{journal}{J. Exp. Teor. Phys.} \textbf{\bibinfo{volume}{86}},
  \bibinfo{pages}{682} (\bibinfo{year}{1998}).

\end{thebibliography}
\end{document}